\documentclass[twocolumn]{aastex631}

\usepackage{color}
\usepackage{url}
\usepackage{hyperref}
\usepackage{xspace}
\usepackage{soul}

\newcommand {\bc}{\begin {center}}
\newcommand {\ec}{\end {center}}
\newcommand {\be}{\begin {equation}}
\newcommand {\ee}{\end {equation}}
\newcommand {\beq}{\begin {eqnarray}}
\newcommand {\eeq}{\end {eqnarray}}

\newcommand {\nustar}{{NuSTAR}\xspace}
\newcommand {\ixpe}{{IXPE}\xspace}
\newcommand {\swift}{{Swift}\xspace}

\newcommand {\vela}{\mbox{Vela~X-1}\xspace}

\shorttitle{X-ray polarimetry of Vela X-1}
\shortauthors{Forsblom S.~V. et al.}

\begin{document}

\title{IXPE observations of the quintessential wind-accreting X-ray pulsar Vela X-1}

\author[0000-0001-9167-2790]{Sofia V. Forsblom}
\affiliation{Department of Physics and Astronomy, FI-20014 University of Turku,  Finland}

\author[0000-0002-0983-0049]{Juri Poutanen}
\affiliation{Department of Physics and Astronomy, FI-20014 University of Turku,  Finland}
\affiliation{Space Research Institute of the Russian Academy of Sciences, Profsoyuznaya Str. 84/32, Moscow 117997, Russia}

\correspondingauthor{Juri Poutanen}
\email{juri.poutanen@utu.fi}

\author[0000-0002-9679-0793]{Sergey~S.~Tsygankov}
\affiliation{Department of Physics and Astronomy, FI-20014 University of Turku,  Finland}
\affiliation{Space Research Institute of the Russian Academy of Sciences, Profsoyuznaya Str. 84/32, Moscow 117997, Russia}

\author[0000-0002-4576-9337]{Matteo Bachetti}
\affiliation{INAF Osservatorio Astronomico di Cagliari, Via della Scienza 5, 09047 Selargius (CA), Italy}

\author[0000-0003-0331-3259]{Alessandro {Di Marco}}
\affiliation{INAF Istituto di Astrofisica e Planetologia Spaziali, Via del Fosso del Cavaliere 100, 00133 Roma, Italy}

\author[0000-0001-8162-1105]{Victor Doroshenko}
\affiliation{Institut f\"ur Astronomie und Astrophysik, Universit\"at T\"ubingen, Sand 1, D-72076 T\"ubingen, Germany}

\author[0000-0001-9739-367X]{Jeremy Heyl}
\affiliation{Department of Physics and Astronomy, University of British Columbia, Vancouver, BC V6T 1Z1, Canada}

\author[0000-0001-8916-4156]{Fabio {La Monaca}}
\affiliation{INAF Istituto di Astrofisica e Planetologia Spaziali, Via del Fosso del Cavaliere 100, 00133 Roma, Italy}

\author[0000-0002-0380-0041]{Christian Malacaria}
\affiliation{International Space Science Institute, Hallerstrasse 
6, 3012 Bern, Switzerland}

\author[0000-0002-6492-1293]{Herman L. Marshall}
\affiliation{MIT Kavli Institute for Astrophysics and Space Research, Massachusetts Institute of Technology, 77 Massachusetts Avenue, Cambridge, MA 02139, USA}

% ...

\collaboration{100}{(IXPE Collaboration)}



\begin{abstract}
The radiation from accreting X-ray pulsars was expected to be highly polarized, with some estimates for the polarization degree of up to $80\%$.
However, phase-resolved and energy-resolved polarimetry of X-ray pulsars is required in order to test different models and to shed light on the emission processes and the geometry of the emission region.
Here we present the first results of the observations of the accreting X-ray pulsar \vela performed with the \textit{Imaging X-ray Polarimetry Explorer} (IXPE). 
\vela is considered to be the archetypal example of a wind-accreting high-mass X-ray binary system, consisting of a highly magnetized neutron star accreting matter from its supergiant stellar companion.
The spectro-polarimetric analysis of the phase-averaged data for \vela reveals a polarization degree (PD) of 2.3$\pm$0.4\% at the polarization angle (PA) of $-47\fdg3\pm5\fdg4$.
A low PD is consistent with the results obtained for other X-ray pulsars and is likely related to the inverse temperature structure of the neutron star atmosphere.
The energy-resolved analysis shows the PD above 5~keV reaching 6--10\%, and a $\sim90\degr$ difference in the PA compared to the data in the 2--3 keV range. 
The phase-resolved spectro-polarimetric analysis finds a PD  in the range 0--9\% with the PA varying  between $-80\degr$ and~$40\degr$.
\end{abstract}



\section{Introduction} 
\label{sec:intro}

Accreting X-ray pulsars (XRPs) harbor some of the strongest magnetic fields in the entire Universe, which can be as large as several times $10^{12}$ G.
The strong magnetic field channels accreting matter onto the polar regions at the neutron star surface, where it produces hot-spots that are bright in the X-rays; these spots rotate in and out of the observer's line-of-sight, resulting in the appearance of pulsed X-ray emission. Highly magnetized XRPs represent unique laboratories and much information is embedded in the interplay between the immense magnetic fields and the accretion flow.
Observations of emission from accreting XRPs therefore constitute a substantial area of interest for theoretical models of matter interactions with ultrastrong magnetic fields, which cannot be replicated in terrestrial laboratories \citep[see][for a recent review]{2022arXiv220414185M}.

The magnetic field of the neutron star is the main cause for the polarized X-ray emission from accreting XRPs.
The scattering of photons in a highly magnetized plasma is expected to result in a large degree of polarization of the emerging X-ray emission, up to 80\% \citep{1988ApJ...324.1056M,2021MNRAS.501..109C}.
\cite{1988ApJ...324.1056M} showed that linear X-ray polarization is strongly dependent on the geometry of the emission region, and also that it varies with energy and pulse phase.
X-ray polarimetric observations of accreting XRPs can therefore be used to probe the geometry of the emission region.
The phase-resolved polarimetry can be used as a tool to constrain the viewing geometry and to distinguish between the models of their radiation.
 
A large window of opportunity to achieve significant observations of polarized X-ray emission opened up recently, thanks to the launch of the \textit{Imaging X-ray Polarimetry Explorer}   \citep[\ixpe,][]{Weisskopf2022}.
\ixpe is the first imaging X-ray polarimetric mission. 
For observations of XRPs, the strategy is to detect pulsations in the polarized emission, as well as to measure the polarization degree (PD) and polarization angle (PA) as a function of the pulse phase, which allows for the system geometry to be determined.

\vela (associated with the Uhuru source 4U 0900--40) is a high-mass X-ray binary (HMXB) discovered as one of the first X-ray sources at the early years of X-ray astronomy \citep{1967ApJ...150...57C} and remains one of the best studied objects among neutron star HMXBs. 
It is one of the brightest, persistent XRPs chosen to be observed by \ixpe.
\vela, located at a distance of about 2 kpc \citep{2021A&A...652A..95K}, is often considered the quintessential wind accretor. 
It displays strong X-ray pulsations with a pulse period of $\mathrm{283\;s}$ \citep{1976ApJ...206L..99M},  variations with the orbital period of $\mathrm{8.964\;days}$ \citep{1972ApJ...178L.121U,1995A&A...303..483V}, and eclipses lasting for about two days per orbit. 
The lower limit on the orbital inclination was obtained at $i=73\degr$ \citep{1995A&A...303..483V}.

Persistent wind-accreting XRPs are expected to have a different emission region geometry as opposed to the disk-accreting Be/X-ray binaries.
Additionally, polarization signatures are expected to be introduced by the scattering in the dense asymmetric wind.
The main goal is therefore to study the accretion geometry for wind-accretion and the properties of the dense stellar wind.

\citet{2003A&A...401..313Q} have shown that the separation between the neutron star and its stellar companion, the B0.5Ib supergiant HD 77581 (also known as GP Vel), is only about 1.7 stellar radii and therefore the neutron star is deeply embedded in the stellar wind of its companion star.
The stellar companion has a mass loss rate of $\sim10^{-6}\;M_\odot$~yr$^{-1}$   \citep{1986PASJ...38..547N}.
The average X-ray luminosity of the source is $\sim4\times10^{36}$\,erg\,s$^{-1}$.
The luminosity is, however, strongly variable on all time-scales, varying up to a factor of at least 20--30 \citep{2004ESASP.552..259S, 2008A&A...492..511K}.

Observations of the cyclotron resonance scattering features (CRSFs) in the spectra of XRPs provide a direct measurement of the magnetic field strength in the line-forming region. 
These features were first discovered in \vela by \citet{1992fxra.conf...51K} by utilizing Mir-HEXE data, reporting a fundamental line around 25 keV and a first harmonic close to 50 keV.
Evidence for these features was also given by \citet{1992fxra.conf...23M} and they were further detailed by \citet{1996A&AS..120C.175K}.
Early observations with {RXTE} also confirmed this detection \citep{1997A&A...325..623K}.
More recent observations of \vela by \nustar clearly detected the fundamental line at 25 keV together with a more prominent first harmonic at 55 keV and revealed a positive correlation between the harmonic line energy and the observed flux \citep{2014ApJ...780..133F}.
\citet{2016MNRAS.463..185L} confirmed a flux dependence of the first harmonic line energy and discovered its secular variation with time in the long-term data collected by \swift/BAT \citep[see also][]{2019MNRAS.484.3797J}.

\begin{figure*}
\centering
\includegraphics[width=0.85\linewidth]{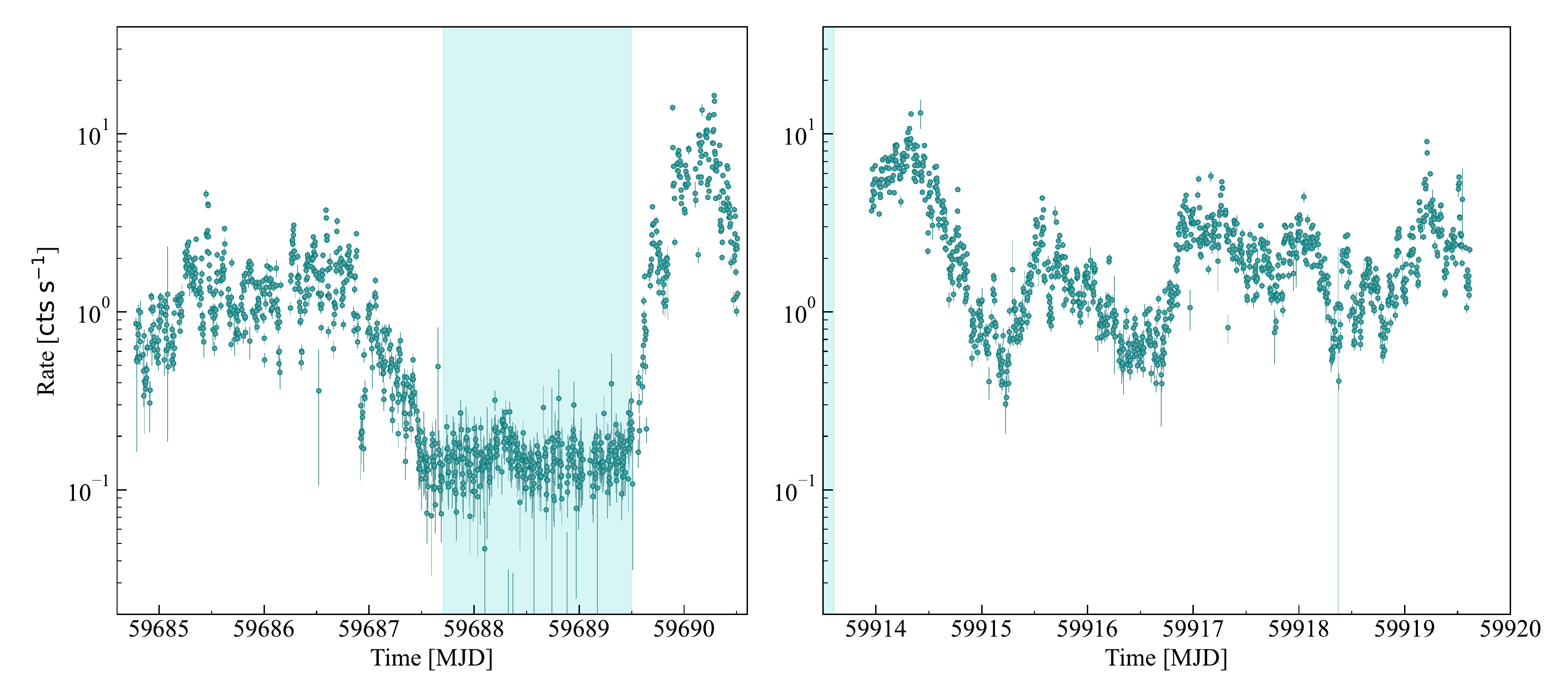}
\caption{Background-corrected light curves of \vela in the 2--8 keV energy bands summed over the three DUs of \ixpe for the first and second observation, shown in the left and right panels, respectively.
The light curve time bin value was set to $\sim$250 s.
The time of the eclipse is marked by the blue-shaded area.
}
 \label{fig:ixpe-lc}
\end{figure*}

In this paper, we present the first results of X-ray polarimetric observations of \vela by \ixpe carried out on two separate occasions during 2022.
In Sect.~\ref{sec:data}, the data used in the paper are described. 
Sect.~\ref{sec:res} is devoted to the description of the results: the analysis of the phase-averaged, phase-resolved, and energy-resolved polarimetric data are given. 
Finally, the discussion and a short summary are presented in Sect.~\ref{sec:sum}.

\section{Data} 
\label{sec:data}

\begin{table}
\centering
\caption{Orbital parameters for \vela adopted from the Fermi Gamma Ray Burst Monitor project (dated 2021 January 30).} 
\begin{tabular}{ccc}
\hline
\hline
 Parameter & Value & Unit\\
\hline
        Orbital period  & 8.9642140 & d  \\
        $T_{\pi/2}$  & 2459115.02085 & JED  \\
        $a_{\rm X}\sin i$  & 113.105 & light-sec  \\
        Longitude of periastron  & 162.33 & deg   \\
        Eccentricity  & 0.0872 &    \\
        Eclipse egress  & 0.12 &    \\
        Eclipse ingress  & 0.92 &    \\
\hline
\end{tabular}
\label{table:GBM-orb-pars}
\end{table}

\ixpe is an observatory launched on December 9, 2021, as a NASA mission in partnership with the Italian space agency (ASI). 
\ixpe consists of three telescope-detector systems which provide imaging polarimetry over a nominal 2--8 keV band at $\sim$ 30\arcsec\ angular resolution (half-power diameter). 
Each one of the three grazing incidence telescopes is comprised of a mirror module assembly (MMA), that focuses the X-rays onto a corresponding focal plane polarization-sensitive gas pixel electron tracking detector unit (DU). 
The detection principle is based on the photoelectric effect. 
All characteristics of each detected photon (sky coordinates, time of arrival, energy, and direction of the photo-electron) are measured simultaneously. A comprehensive description of the observatory, the instruments, and their performance is given by \citet{Soffitta21} and \citet{Weisskopf2022}.

\ixpe observations of \vela were carried out between 2022 April 15--21 and November 30 -- December 6, with the total effective exposure of $\simeq$280~ks and $\simeq$270~ks, respectively.
The data have been processed with the {\sc ixpeobssim} package version 30.2.1 \citep{Baldini2022} using the CalDB released on November 17, 2022.
The position offset correction and energy calibration were applied before the scientific analysis of the data. 
Source photons were collected using a circular region with a radius $R_{\rm src}=70\arcsec$.
The background region was chosen in the form of an annulus with inner and outer radii equal to 2$R_{\rm src}$ and 4$R_{\rm src}$, respectively.
Data from the first observation were cleaned from events due to solar events, which have been identified by comparing the IXPE light curve with the one from the Geostationary Operational Environmental Satellite (GOES), then removing time intervals where the IXPE count rate in the background annular region was higher than the mean background value plus three times the RMS of this count rate. 
The background makes up $\sim$3.6\% and $\sim$2.2\% of the total count rate of the source region in the 2--8 keV energy range for the first and second observation, respectively.

\begin{figure*}
\centering
\includegraphics[width=0.7\linewidth]{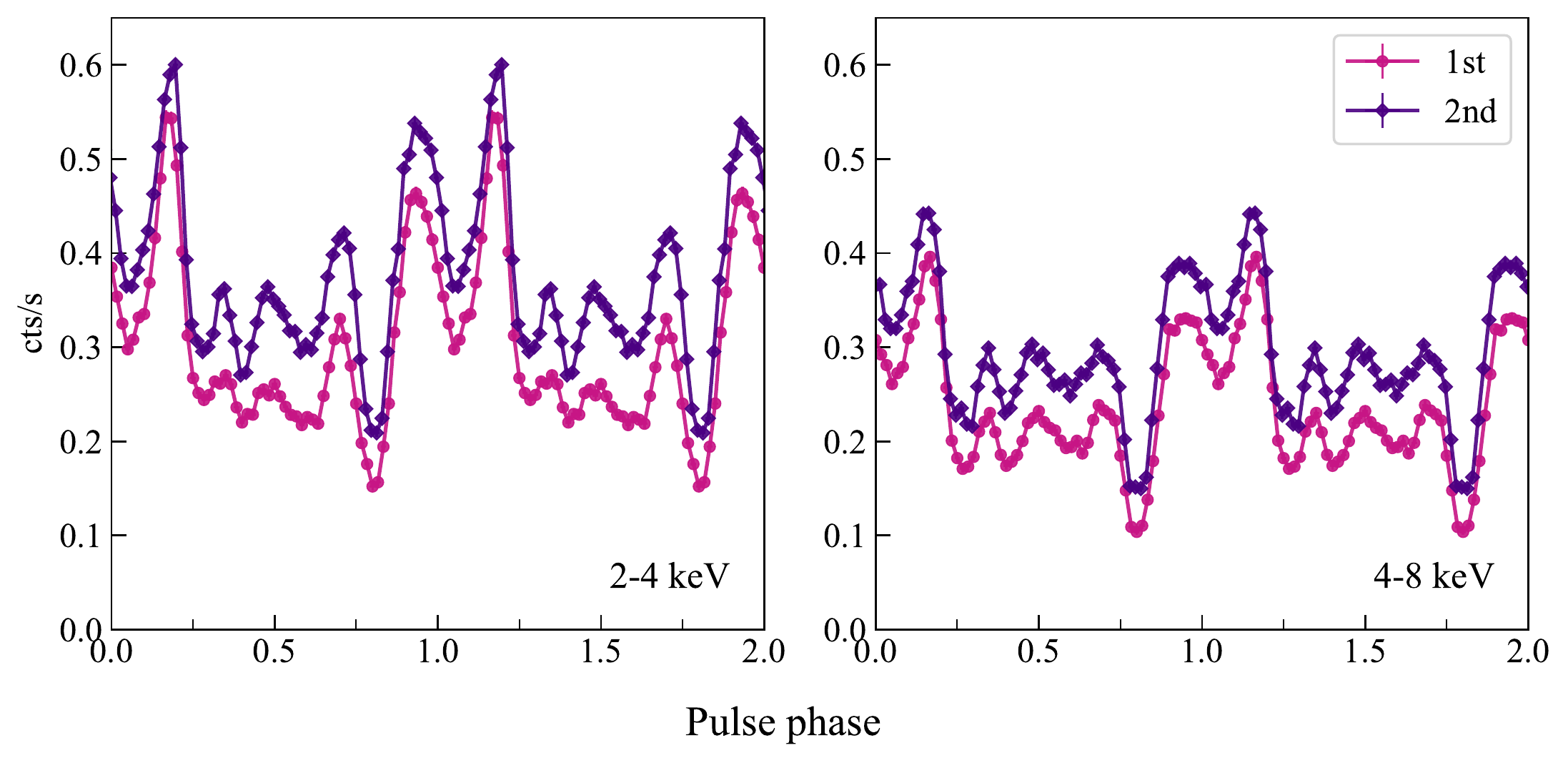}
\caption{Pulse profiles of \vela as seen by \ixpe in two different energy bands for the first and second observations, combined for DU1--3.}
\label{fig:efold-pprof}
\end{figure*}

The \texttt{barycorr} tool from the {\sc ftools} package was used to correct the event arrival times to the barycenter of the Solar System. This was followed by a correction of arrival times as it relates to the effects of binary motion using the orbital parameters obtained by the \textit{Fermi} Gamma Ray Burst Monitor\footnote{\url{https://gammaray.nsstc.nasa.gov/gbm/science/pulsars.html}} for \vela and given in Table~\ref{table:GBM-orb-pars}.

Stokes $I$ energy spectra have been binned to have at least 30 counts per energy channel, and the same energy binning was applied to the energy spectra of Stokes parameters $Q$ and $U$. The energy spectra were fitted in the {\sc xspec} package \citep{Arn96} using $\chi^2$ statistics, using the version 12 instrument response functions.
The reported uncertainties are at the 68.3\% confidence level ($1\sigma$), unless stated otherwise.

\section{Results} 
\label{sec:res}

\subsection{Light curve and pulse profile} 
\label{sec:lc}

The light curves from the first and second observation of Vela X-1 in the 2--8 keV energy range obtained with the \ixpe observatory are shown in Figure~\ref{fig:ixpe-lc}. 
For the first observation of \vela, the observing window can be separated into three parts: pre-eclipse (MJD 59684.7--59687.7), eclipse (MJD 59687.7--59689.5), and post-eclipse (MJD  59689.5--59690.5). The post-eclipse count rate of the source was about one order of magnitude greater than the pre-eclipse source count rate.
During the eclipse, which lasted about two days, the count rate dropped by an order of magnitude compared to the pre-eclipse value.
For the following analysis, only pre- and post-eclipse data were included, i.e., only data outside of the eclipse. 

A spin period of $P_{\rm spin} = 283.488(7)$~s and $P_{\rm spin} = 283.437(6)$~s were measured for \vela for the first and second observation, respectively, using phase connection technique.
The pulsed fraction in the 2--8 energy band, defined as $PF = (F_{\max}-F_{\min})/(F_{\max}+F_{\min})$ where $F_{\max}$ and $F_{\min}$ are the maximum and minimum count rates in the pulse profile, respectively, was determined as $PF = 53.3\pm0.7\%$ for the first observation and as $PF = 48.1\pm0.9\%$ for the second observation. 
The resulting pulse profiles for \vela in two separate energy bands are shown in Figure~\ref{fig:efold-pprof}. 
 
\subsection{Polarimetric analysis} 
\label{sec:ph-ave}

\begin{table}
\centering
\caption{Measurements of the normalized Stokes parameters $q$ and $u$, PD, and PA for the phase-averaged data of \vela in different energy bins using the \texttt{PCUBE} algorithm for the combined data set. 
}
\begin{tabular}{ccccc}
    \hline\hline
    Energy  & $q$ & $u$ & PD & PA \\ 
       (keV)   & (\%) & (\%) & (\%) & (deg) \\ 
    \hline
    2--3 & $\phantom{-}0.5\pm1.2$ & $\phantom{-}3.8\pm1.2$ & 3.9$\pm$1.2 & $\phantom{-}41.4\pm9.1$ \\ 
    3--4 & $\phantom{-}0.2\pm0.8$ & $-1.4\pm0.8$ & 1.4$\pm$0.8 & $-40.7\pm16.1$  \\ 
    4--5 & $-0.1\pm0.8$ & $-2.7\pm0.8$ & 2.7$\pm$0.8 & $-46.4\pm8.8$ \\ 
    5--6 & $-1.2\pm1.0$ & $-5.6\pm1.0$ & 5.7$\pm$1.0 & $-50.9\pm4.8$ \\ 
    6--7 & $-0.3\pm1.2$ & $-4.1\pm1.2$ & 4.1$\pm$1.2 & $-47.3\pm8.3$ \\ 
    7--8 & $-2.7\pm2.0$ & $-9.4\pm2.0$ & 9.7$\pm$2.0 & $-53.1\pm6.0$ \\ 
    \hline
     2--8 & $-0.6\pm0.5$& $-3.7\pm0.5$ & 3.7$\pm$0.5 & $-49.9\pm4.1$ \\ 
    \hline
    \end{tabular}
    \label{table:ebins}
\end{table}

\begin{figure}
\centering
\includegraphics[width=0.84\linewidth]{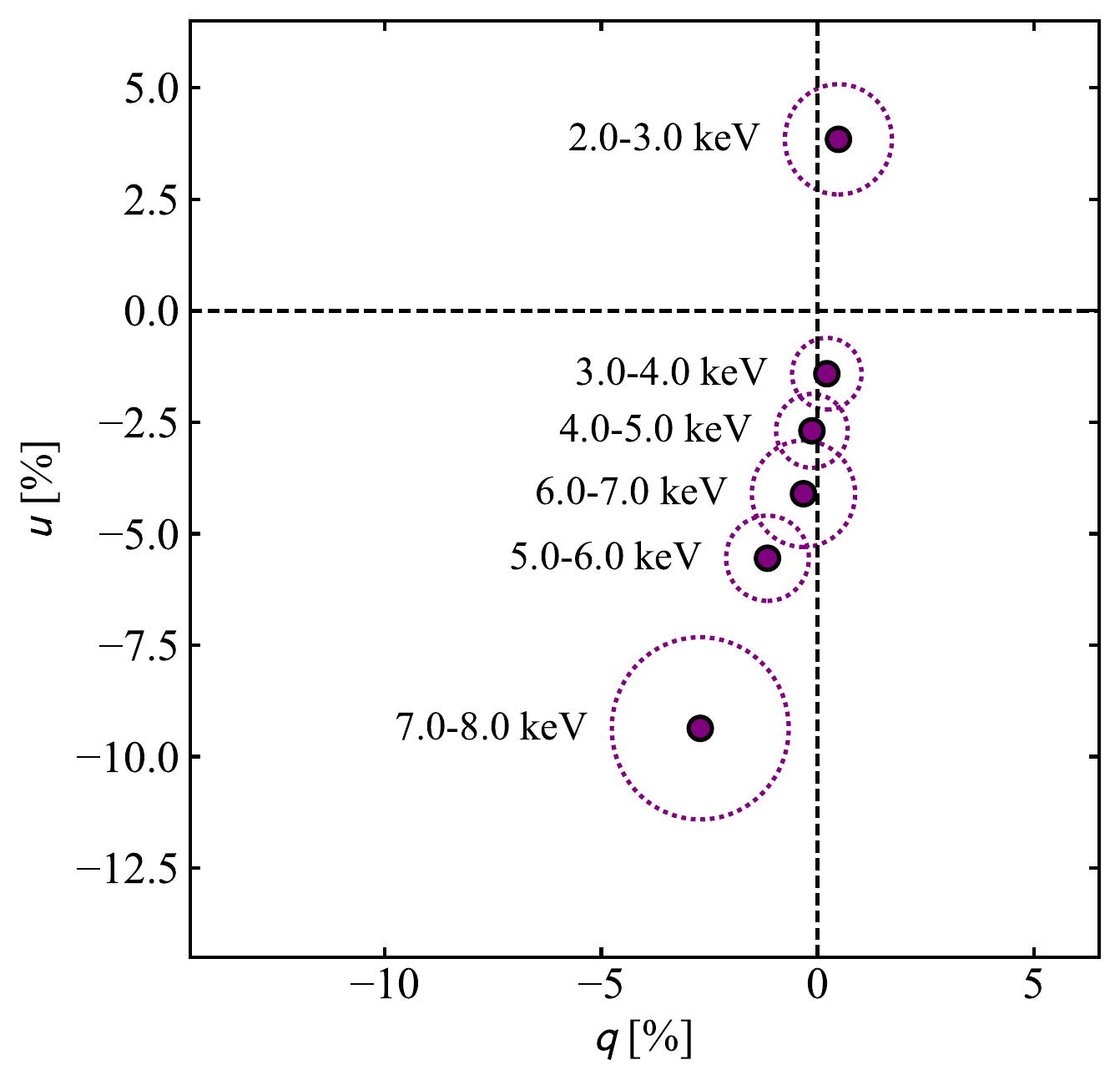}
\caption{Energy dependence of the normalized Stokes parameters $q$ and $u$ for the phase-averaged data of the combined data set, obtained with the \texttt{PCUBE} algorithm. } 
\label{fig:phase-ave-PCUBE}
\end{figure}

\begin{figure}
\centering
\includegraphics[width=0.9\linewidth]{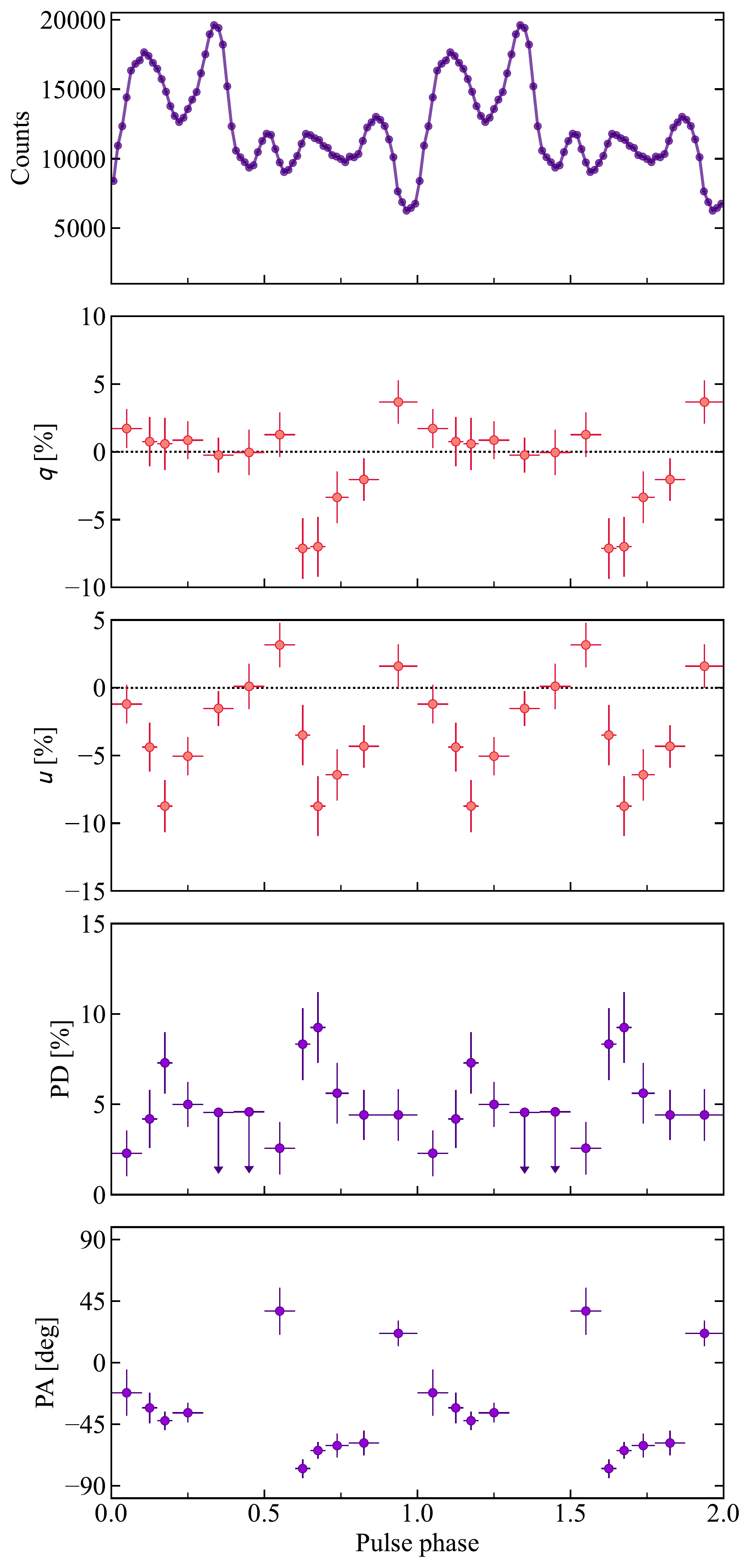}
\caption{Phase-resolved analysis of \vela for the combined data set in the 2--7 keV range, combining data from all DUs.  
(a) Pulse profile.
Panels (b) and (c) display the dependence of the Stokes $q$ and $u$ parameters,  respectively, on the pulse phase, obtained for the phase-resolved polarimetric analysis utilizing the \texttt{PCUBE} algorithm.
Panels (d) and (e) show the dependence of the PD and PA, respectively, on the pulse phase, obtained from the phase-resolved spectro-polarimetric analysis using {\sc xspec}. Upper limits (arrows) to the PD are at 99.73\% (3$\sigma$) confidence level and are computed using a $\chi^2$ with one degree of freedom.  
}
 \label{fig:phase-res-pcube-xspec}
\end{figure}

First, the analysis of the polarimetric properties of \vela was carried out by using the \texttt{PCUBE} algorithm (\texttt{xpbin} tool) in the {\sc ixpeobssim} package, which is implemented according to the formalism by \citet{2015-Kislat}. 
The unweighted analysis has been used. 
We compute the normalized Stokes parameters, $q=Q/I$ and $u=U/I$, and the polarization degree using the formula PD=$\sqrt{q^2+u^2}$ and ignoring the bias at low signal-to-noise ratios  \citep{Serkowski1958,Simmons1985,Maier2014,Mikhalev2018} and the PA=$\frac{1}{2}\arctan (u/q)$ (measured counterclockwise on the sky from north to east).

In the entire IXPE energy band (2--8 keV), the average PD and PA are found to be 3.9$\pm$0.9\% and $-51\fdg5\pm6\fdg5$, respectively, for the first observation.
For the second observation, the average PD and PA are found to be 3.7$\pm$0.7\% and $-48\fdg9\pm5\fdg2$, respectively.

Considering the similarities between the first and second observations, the data were combined into one single set of data in order to increase the statistics and further study the polarization properties of \vela.
In order to correctly phase-tag each event, the phase difference between the pulse profiles from the first and second observations was determined from cross-correlation (using the implementation provided by the Python library NumPy).
Using the \texttt{PCUBE} algorithm (\texttt{xpbin} tool) in the {\sc ixpeobssim} package, the average PD and PA are found to be 3.7$\pm$0.5\% and $-49\fdg9\pm4\fdg1$, respectively, in the entire IXPE energy band.

We then studied the energy dependence of polarization by dividing the data into six energy bins.
The PD is above the minimum detectable polarization at a 99\% confidence level, $\mathrm{MDP_{99}}$ \citep{2010SPIE.7732E..0EW}, in all of the energy bins, except for the 3--4 keV bin, where the PD is below the $\mathrm{MDP_{99}}$ and in this case, the PA is not well constrained  (Table~\ref{table:ebins} and Figure~\ref{fig:phase-ave-PCUBE}).
The energy-resolved analysis shows that at higher energies (above 5 keV) the PD reaches 6--10\% with the PA differing by $\sim$90\degr\ from that below 3 keV.

\begin{figure*}
\centering
 \includegraphics[width=0.30\linewidth]{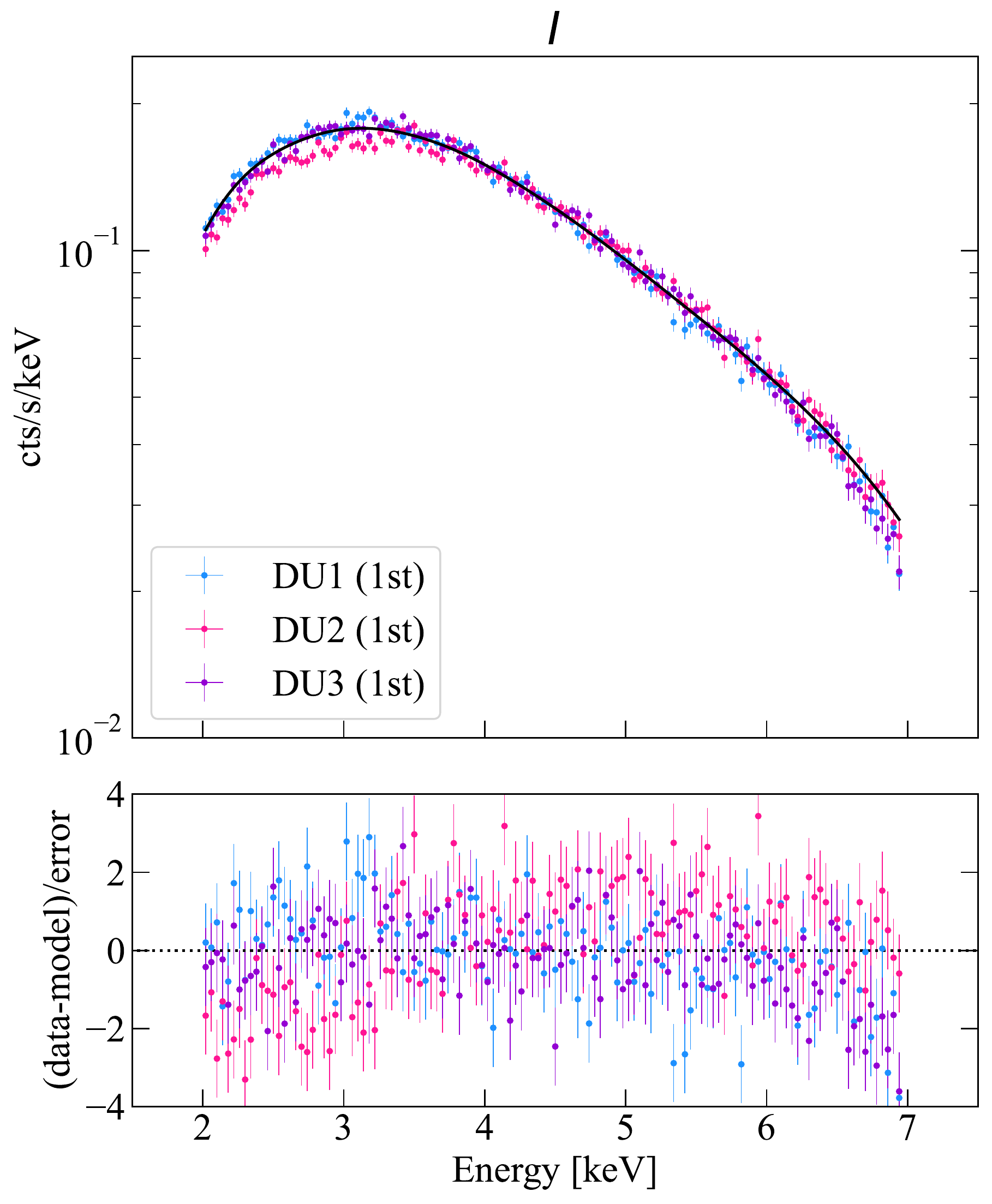}
 \includegraphics[width=0.30\linewidth]{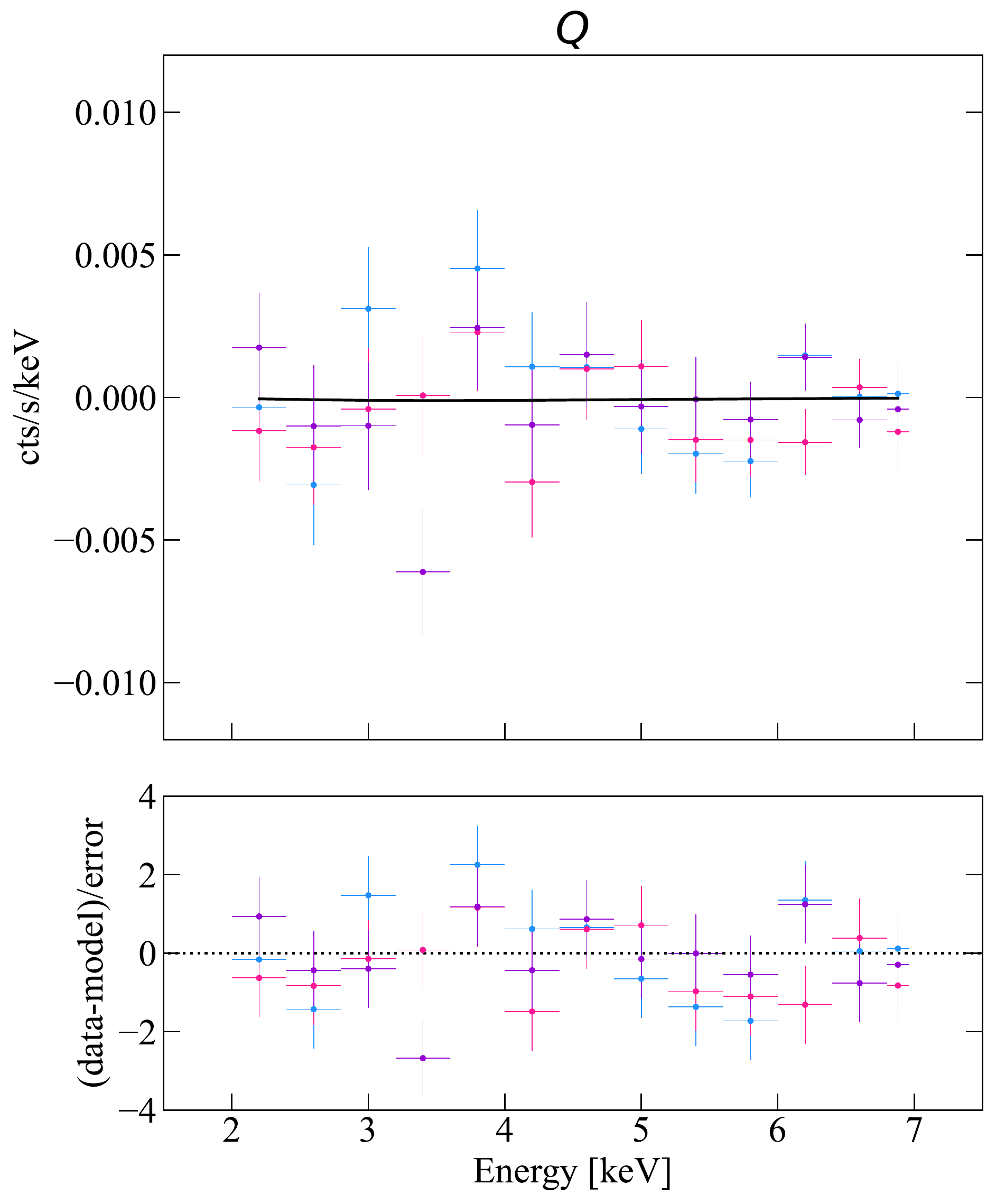}
 \includegraphics[width=0.30\linewidth]{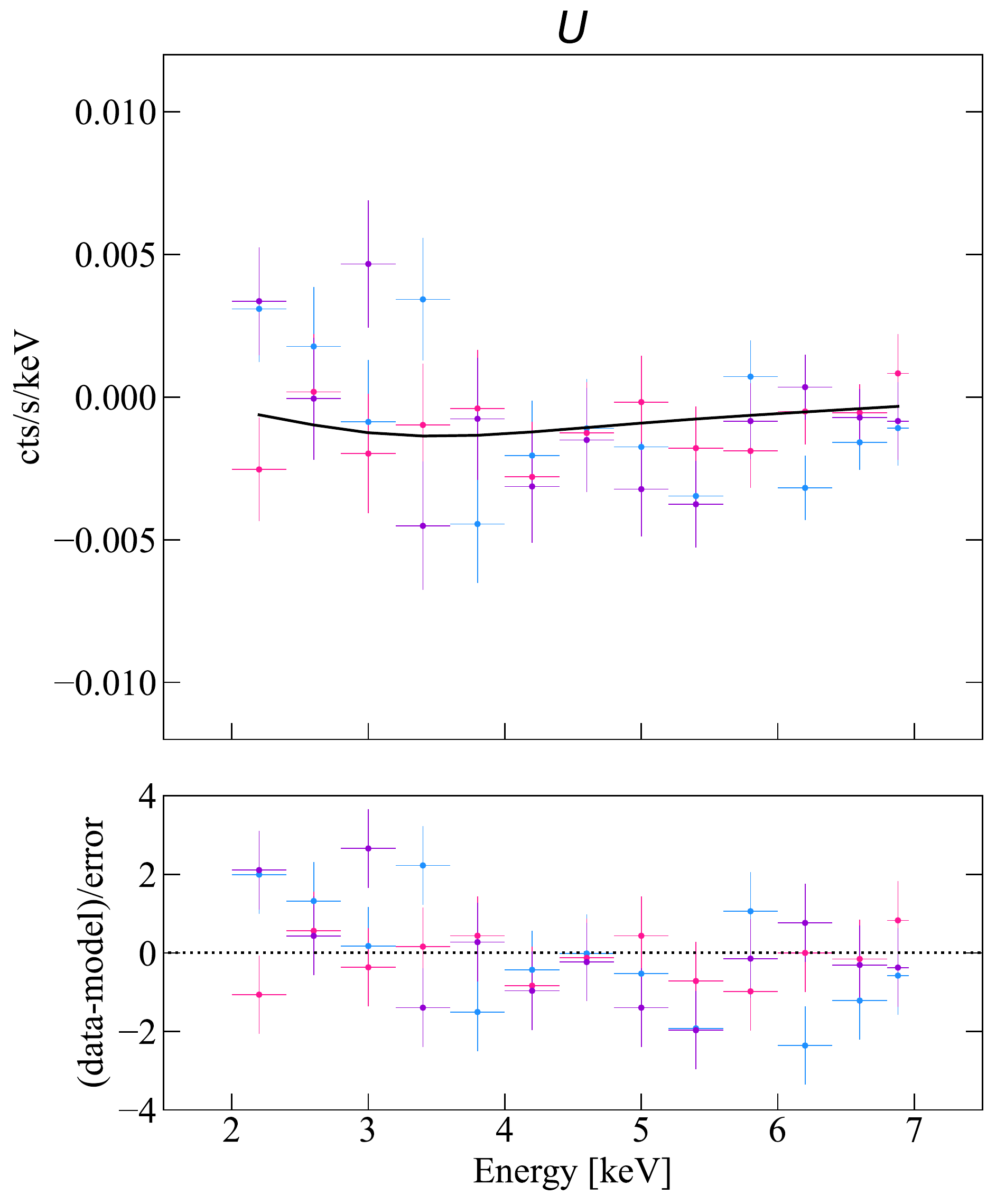}
 \includegraphics[width=0.30\linewidth]{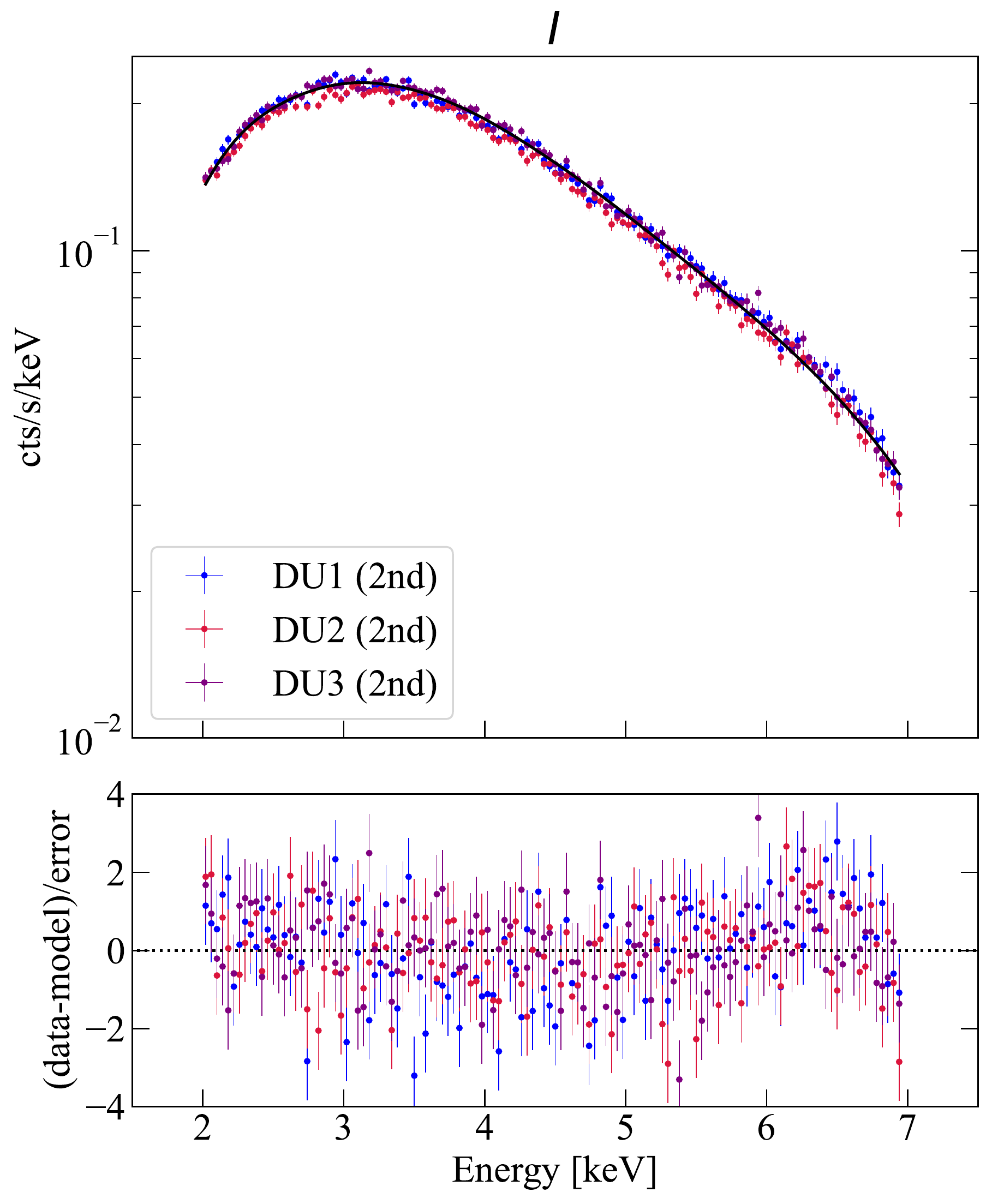}
 \includegraphics[width=0.30\linewidth]{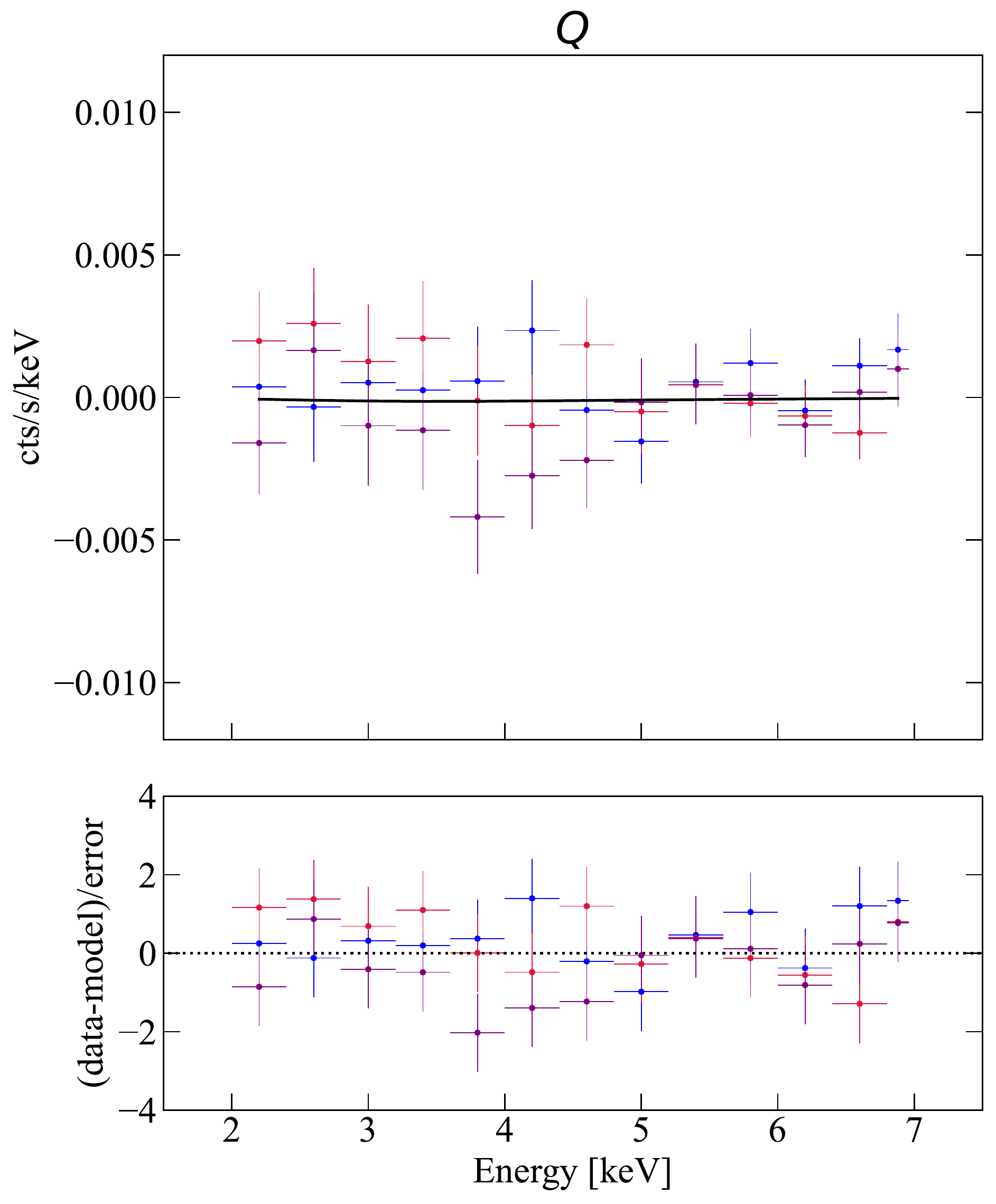}
 \includegraphics[width=0.30\linewidth]{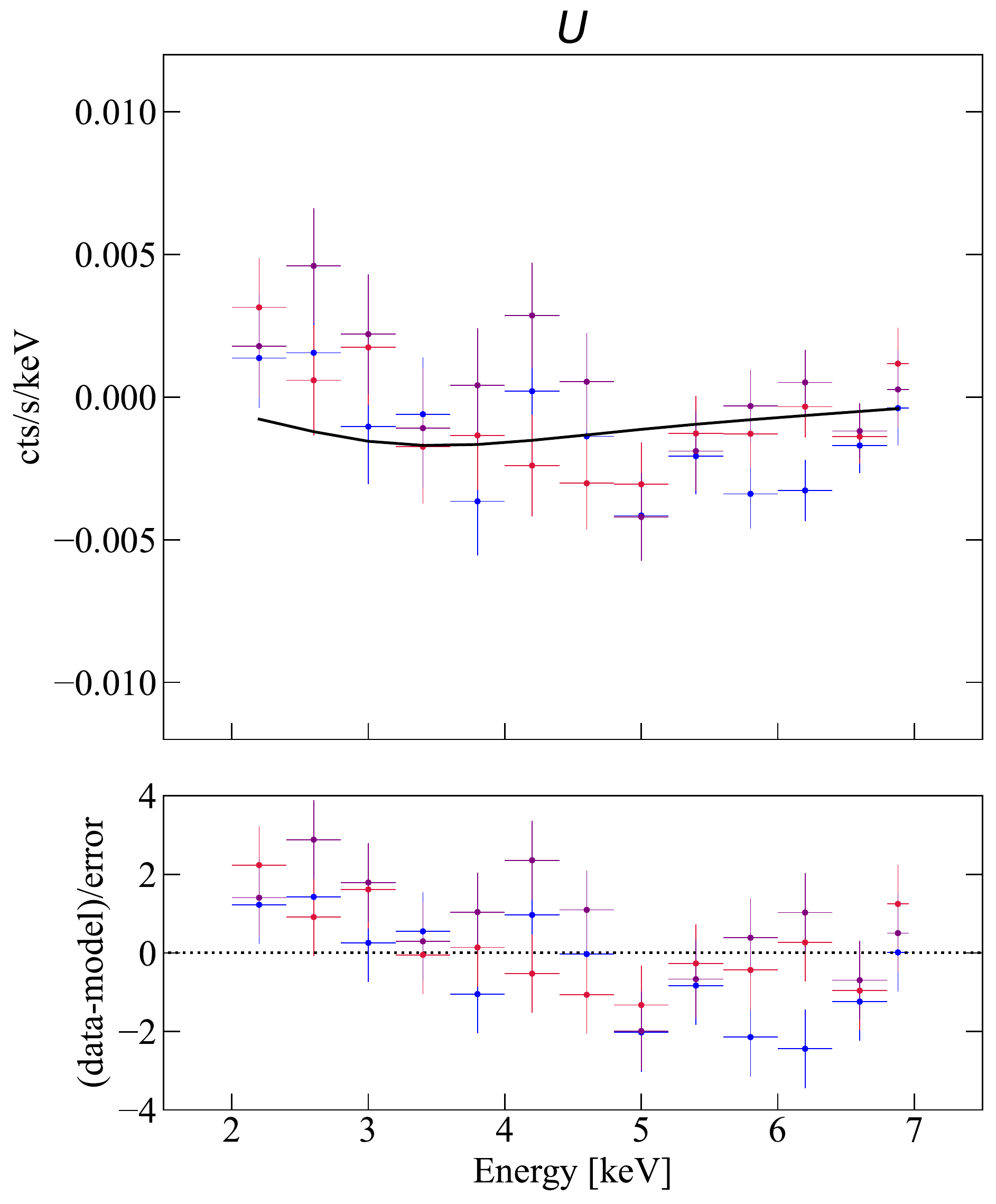}
\caption{Stokes $I$, $Q$, and $U$ energy distributions for the combined data set of \vela with the best-fit model superimposed for IXPE's DUs (upper panels).
The residuals between the best-fit model and the data are shown in the lower panels.
Upper and lower plots correspond to the first and second observations, respectively, and are shown separately for clarity.
}
 \label{fig:xspec}
\end{figure*}

\begin{table*}
\centering
\caption{Pulse phase dependence of the normalized Stokes $q$ and $u$ parameters for the combined data set from the polarimetric analysis (2--7 keV) using the \texttt{PCUBE} algorithm and the spectral parameters, PD and PA obtained by the spectro-polarimetric analysis.
An upper limit to the PD at 99.73\% (3$\sigma$) confidence level is computed using a $\chi^2$ with one degree of freedom.  
}
\begin{tabular}{cccccccccc}
    \hline\hline
    Phase & $q$ & $u$ & $N_{\mathrm{H}}$ & $N_{\mathrm{H,tbpcf}}$ & $f_{\mathrm{cov}}$ & Photon index & PD & PA & $\chi^2$/d.o.f. \\
           & (\%) & (\%) & ($10^{22}\mathrm{\;cm^{-2}}$) & ($10^{22}\mathrm{\;cm^{-2}}$) &  &  & (\%) & (deg) &  \\
    \hline
    0.000--0.100 & $\phantom{-}1.7\pm1.4$ & $-1.2\pm1.4$ & $6.0_{-0.8}^{+0.7}$ & $23.0_{-2.2}^{+1.8}$ & $0.82\pm0.02$ & $1.55_{-0.16}^{+0.13}$ & $2.3\pm1.2$ & $-22.0\pm16.9$ & 2160/2152\\
    0.100--0.150 & $\phantom{-}0.8\pm1.8$ & $-4.4\pm1.8$ & $4.4_{-1.4}^{+1.0}$ & $18.5_{-2.0}^{+2.1}$ & $0.83_{-0.03}^{+0.04}$ & $1.26\pm0.16$ & $4.2\pm1.6$ & $-33.0\pm11.3$ & 2162/2029\\
    0.150--0.200 & $\phantom{-}0.6\pm1.9$ & $-8.7\pm1.9$ & $4.9_{-2.1}^{+1.2}$ & $18.4_{-3.4}^{+3.2}$ & $0.77_{-0.04}^{+0.07}$ & $0.96_{-0.22}^{+0.19}$ & $7.3\pm1.7$ & $-42.4\pm6.8$ & 2081/2011\\
    0.200--0.300 & $\phantom{-}0.9\pm1.4$ & $-5.1\pm1.4$ & $3.8_{-0.9}^{+0.8}$ & $19.9_{-1.9}^{+2.1}$ & $0.80\pm0.02$ & $0.75_{-0.13}^{+0.14}$ & $5.0\pm1.2$ & $-36.6\pm7.2$ & 2284/2179\\
    0.300--0.400 & $-0.2\pm1.3$ & $-1.5\pm1.3$ & $4.0_{-1.1}^{+0.9}$ & $16.5_{-1.4}^{+1.7}$ & $0.81_{-0.03}^{+0.04}$ & $1.11\pm0.11$ & $<4.6$ & --- & 2304/2185\\
    0.400--0.500 & $\phantom{-}0.0\pm1.7$ & $\phantom{-}0.1\pm1.7$ & $4.2_{-1.2}^{+0.8}$ & $20.5_{-2.9}^{+2.5}$ & $0.78\pm0.03$ & $1.08\pm0.2$ & $<4.6$ & --- & 2057/2086\\
    0.500--0.600 & $\phantom{-}1.3\pm1.6$ & $\phantom{-}3.2\pm1.6$ & $2.1_{-1.2}^{+0.9}$ & $17.7_{-1.7}^{+1.7}$ & $0.83\pm0.03$ & $0.63\pm0.13$ & $2.6\pm1.5$ & $37.7\pm17.2$ & 2230/2116\\
    0.600--0.650 & $-7.1\pm2.2$ & $-3.5\pm2.2$ & $3.8_{-1.5}^{+1.0}$ & $21.3_{-3.0}^{+2.6}$ & $0.82\pm0.03$ & $0.98_{-0.22}^{+0.19}$ & $8.3\pm2.0$ & $-77.3\pm6.9$ & 1885/1858\\
    0.650--0.700 & $-7.0\pm2.2$ & $-8.7\pm2.2$ & $2.5_{-1.8}^{+1.2}$ & $18.2_{-2.1}^{+2.4}$ & $0.84\pm0.04$ & $0.73\pm0.18$ & $9.2\pm2.0$ & $-64.2\pm6.1$ & 1934/1912\\
    0.700--0.775 & $-3.4\pm1.9$ & $-6.4\pm1.9$ & $0.6_{-0.6}^{+1.5}$ & $15.3_{-1.2}^{+2.0}$ & $0.85_{-0.05}^{+0.02}$ & $0.30_{-0.13}^{+0.15}$ & $5.6\pm1.6$ & $-60.6\pm8.7$ & 2025/2038\\
    0.775--0.875 & $-2.0\pm1.6$ & $-4.3\pm1.6$ & $4.0_{-1.0}^{+0.7}$ & $20.6_{-2.2}^{+2.0}$ & $0.81\pm0.02$ & $1.05_{-0.15}^{+0.14}$ & $4.4\pm1.4$ & $-58.7\pm9.2$ & 2117/2134\\
    0.875--1.000 & $\phantom{-}3.7\pm1.6$ & $\phantom{-}1.6\pm1.6$ & $3.1_{-2.1}^{+1.2}$ & $14.9_{-1.7}^{+1.9}$ & $0.81_{-0.05}^{+0.07}$ & $0.95_{-0.15}^{+0.13}$ & $4.4\pm1.4$ & $21.4\pm9.5$ & 2271/2092\\
    \hline
    \end{tabular}
    \label{table:phbins}
\end{table*}

Next, a phase-resolved polarimetric analysis was performed utilizing the \texttt{PCUBE}-algorithm. 
The results in the 2--7 keV energy band are given in Table~\ref{table:phbins} and are shown in Figure~\ref{fig:phase-res-pcube-xspec} for the combined data set.

Then, the spectro-polarimetric analysis was performed according to the following steps. 
Source $I$, $Q$, and $U$ Stokes spectra were produced via the \texttt{xpbin} tool’s \texttt{PHA1}, \texttt{PHA1Q}, and \texttt{PHA1U} algorithms, producing a full data set comprised of nine spectra per observation, three for each DU. As the background region is dominated by source events, background subtraction is not applied \citep{2023-IXPE-background}.
Here we also used unweighted analysis. 
The {\sc xspec} package (version 12.12.1) \citep{Arn96}, which is a part of the standard high-energy astrophysics software suite HEASOFT, was used to study polarization as a function of energy. 
All 18 spectra were fitted simultaneously in {\sc xspec}.

There are several phenomenological spectral models used to describe the spectral continuum of \vela.
However, it is well known that except for a soft excess below 3 keV, the X-ray emission below 10 keV can be well described by a simple absorbed power law with an iron line at 6.4 keV.
Due to the restricted energy range covered by IXPE and the energy resolution of the instrument \citep{Weisskopf2022}, we used a simple model consisting of a power law affected by interstellar absorption (model \texttt{tbabs} with the abundances from \citealt{Wilms2000}) combined with the \texttt{polconst} polarization model, which assumes energy-independent PD and PA.
In order to account for the soft excess below 3 keV, a partial covering fraction absorption (model \texttt{tbpcf}) was introduced as well, which applies an added column density to a fraction of the power law.
The re-normalization constant, \texttt{const}, was used to account for the possibility of discrepancies between the different DUs, and for DU1 it was fixed to unity. The final model 
\begin{eqnarray*} 
\texttt{ tbabs$\times$tbpcf$\times$polconst$\times$powerlaw$\times$const}
\label{eq:fit}
\end{eqnarray*}
was subsequently applied to both the phase-averaged and phase-resolved data.
The spectral analysis was confined to the 2--7 keV energy band, ignoring photons above 7 keV due to remaining calibration uncertainties.

\begin{table}
\centering
\caption{Spectral parameters for the best-fit model obtained from the phase-averaged spectro-polarimetric analysis of the combined data set.}
\begin{tabular}{ccc}
\hline
\hline
 Parameter & Value & Unit\\
\hline
        $N_{\mathrm{H}}$   & $3.8\pm0.3$ & $10^{22}\mathrm{\;cm^{-2}}$   \\
        $N_{\mathrm{H,tbpcf}}$   & $18.5_{-0.6}^{+0.7}$ & $10^{22}\mathrm{\;cm^{-2}}$   \\
        $f_{\mathrm{cov}}$   & $0.80\pm0.01$ &    \\
        $\mathrm{const_{DU2}}$    & $0.964\pm0.003$ & \\
        $\mathrm{const_{DU3}}$    & $0.923\pm0.002$ & \\
        Photon index   & $0.93\pm0.04$ & \\
        PD & $2.3\pm0.4$ & \% \\
        PA & $-47.3\pm5.4$ & deg \\
        $\mathrm{Flux_{2-8\;keV}}$ & $7.93_{-0.06}^{+0.02}$ & $\mathrm{10^{-10}\;erg\;cm^{-2}\;s^{-1}}$ \\
        $\mathrm{Luminosity_{2-8\;keV}}$ & $3.8\times10^{35}$ & $\mathrm{erg\;s^{-1}}$ at $d=2.0$ kpc \\
        $\chi^2$ (d.o.f.) & 2640.19 (2222) & \\ 
        \hline
\end{tabular}
\label{table:best-fit}
\end{table}

\begin{figure}
\centering
\includegraphics[width=0.80\linewidth]{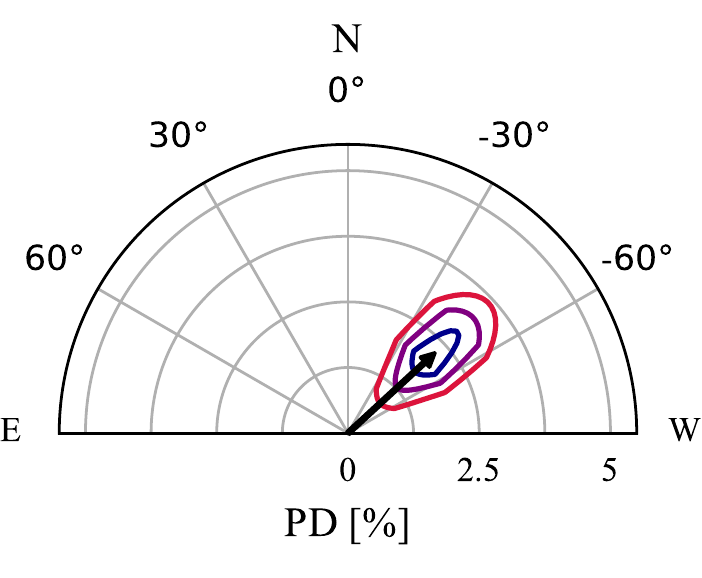}
\caption{Polarization vector of \vela from the results of the phase-averaged spectro-polarimetric analysis of the combined data set. Contours at 68.3\%, 95.45\% and 99.73\% confidence, are shown in blue, purple, and red, respectively. 
}
\label{fig:phase-ave-contour}
\end{figure}

\begin{figure*}
\centering
\includegraphics[width=0.30\linewidth]{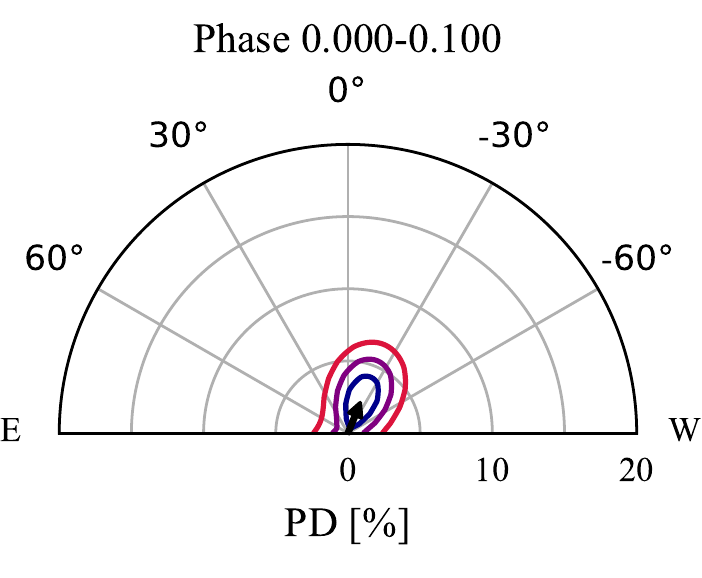}
\includegraphics[width=0.30\linewidth]{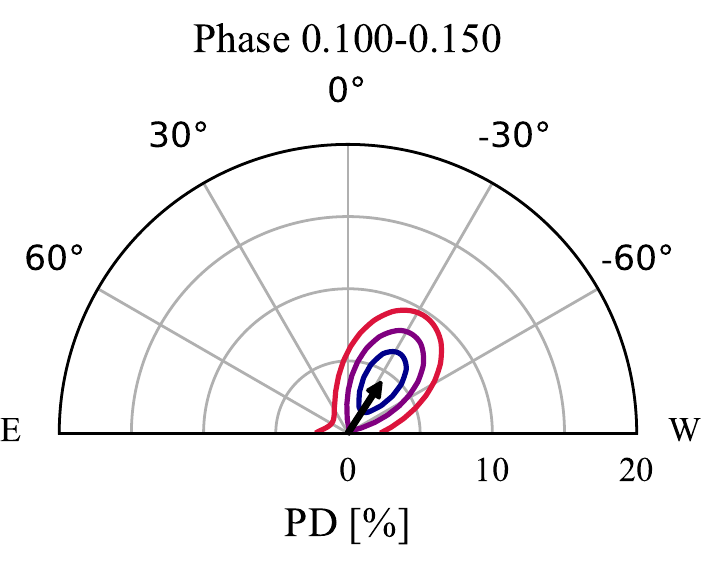}
\includegraphics[width=0.30\linewidth]{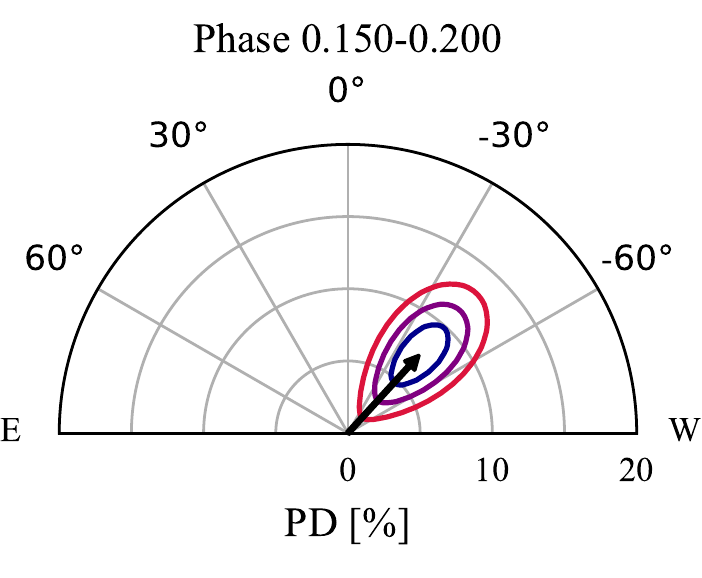}
\includegraphics[width=0.30\linewidth]{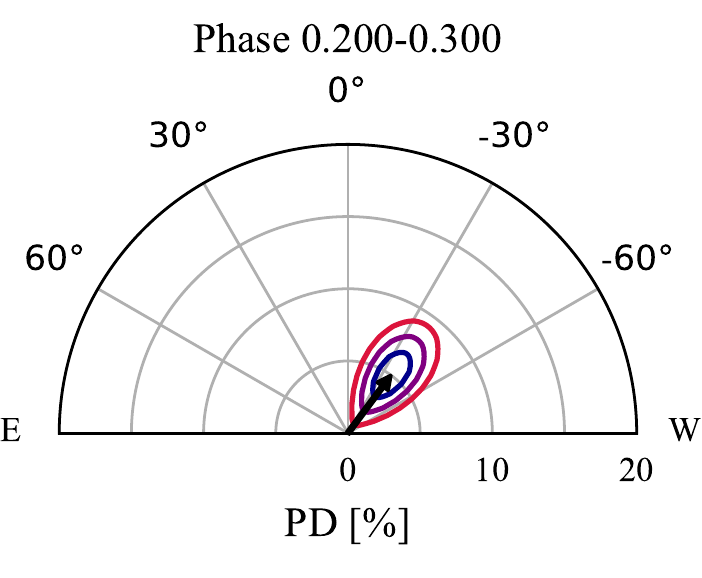}
\includegraphics[width=0.30\linewidth]{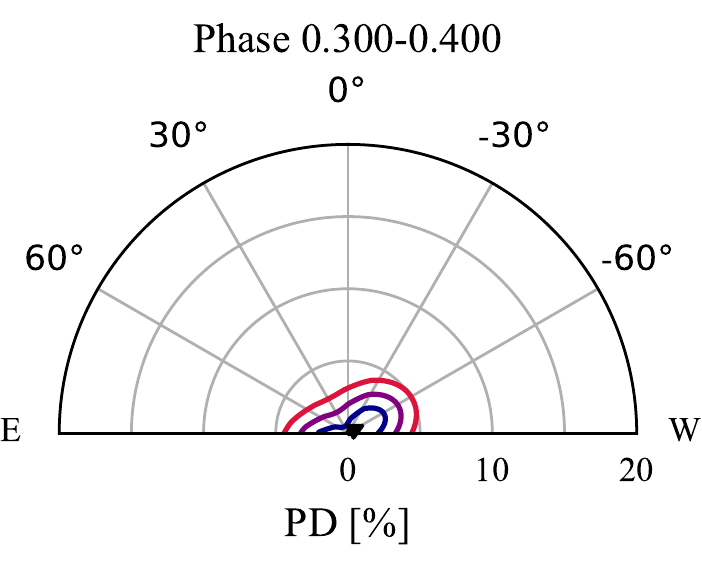}
\includegraphics[width=0.30\linewidth]{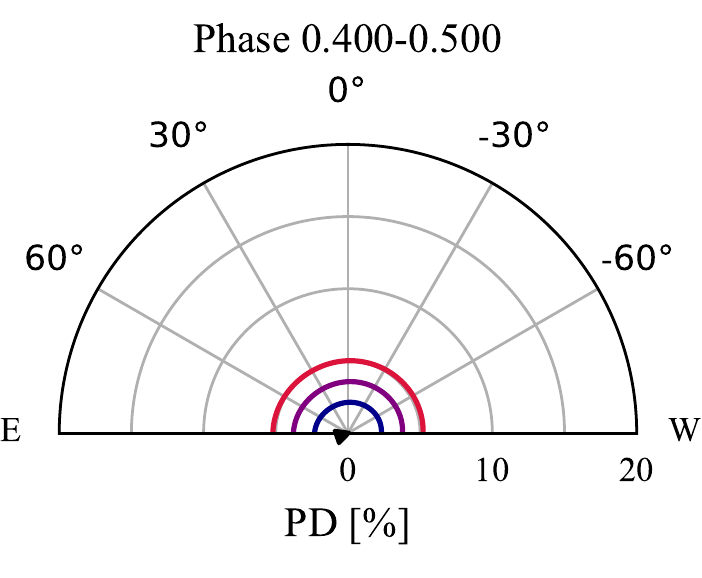}
\includegraphics[width=0.30\linewidth]{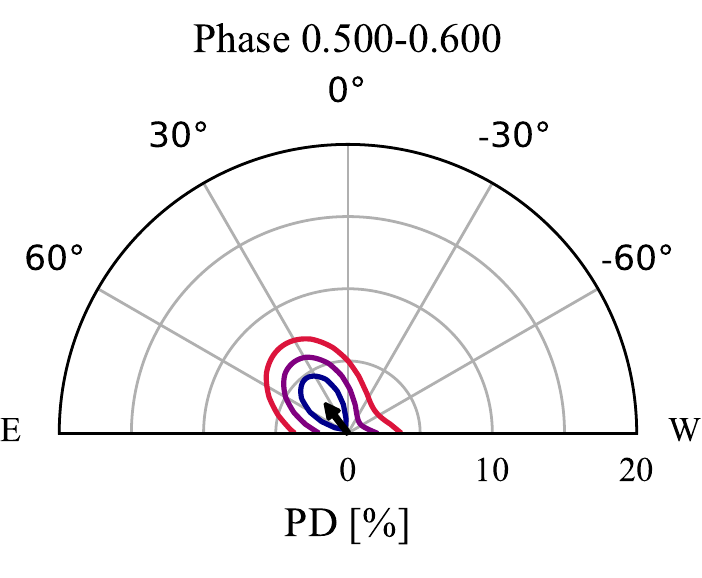}
\includegraphics[width=0.30\linewidth]{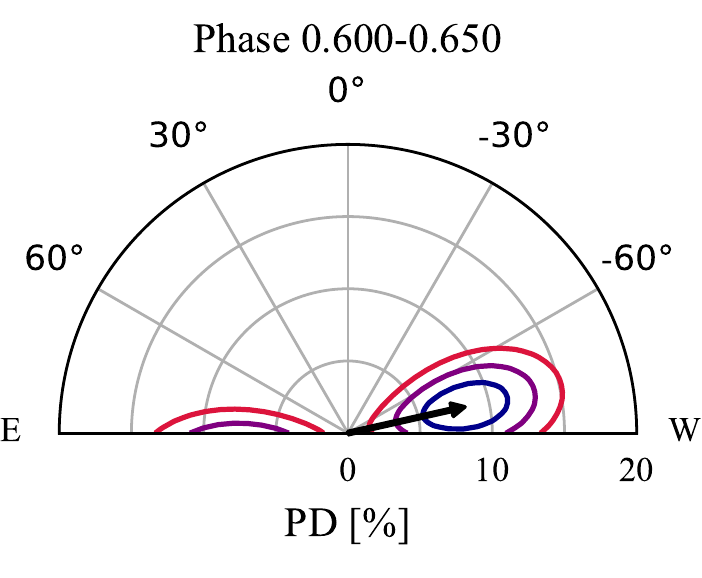}
\includegraphics[width=0.30\linewidth]{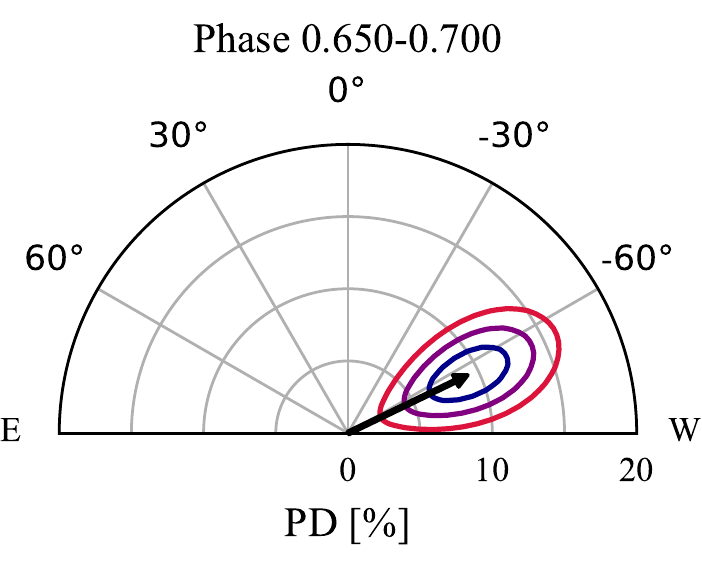}
\includegraphics[width=0.30\linewidth]{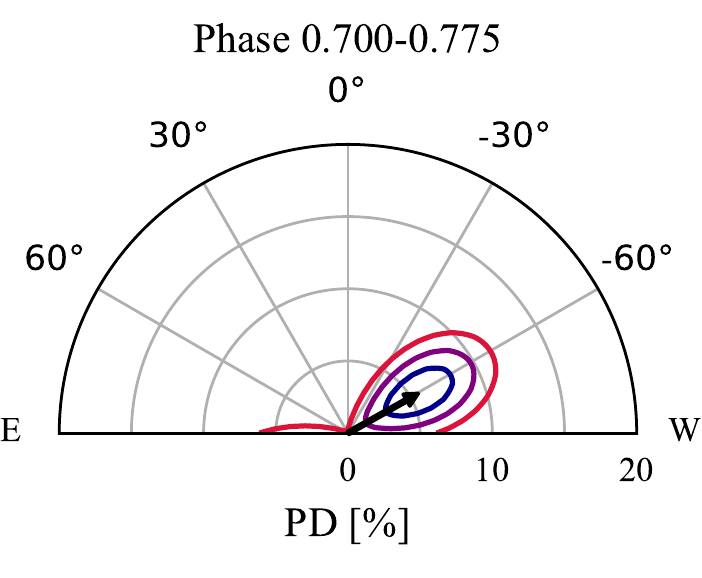}
\includegraphics[width=0.30\linewidth]{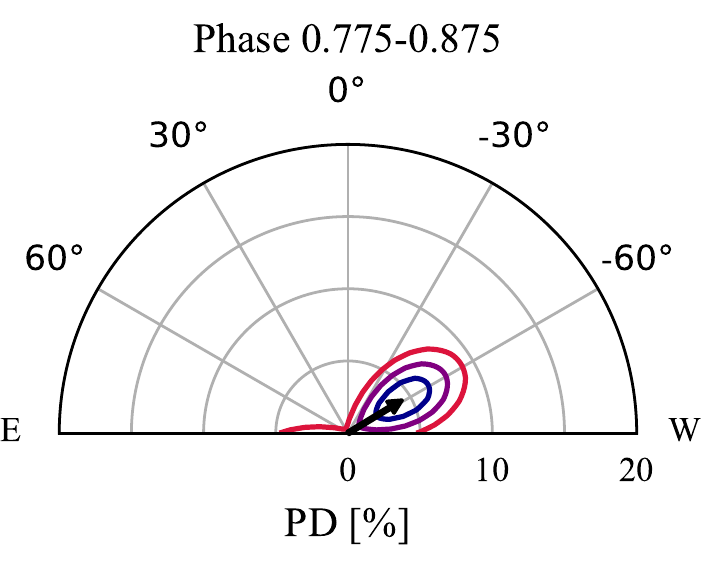}
\includegraphics[width=0.30\linewidth]{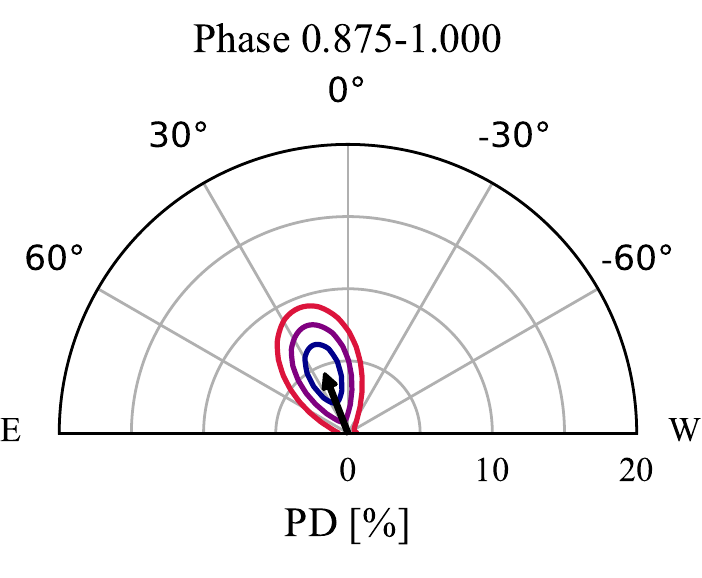}
\caption{Polarization vectors of \vela from the results of the phase-resolved spectro-polarimetric analysis of the combined data set. Contours at 68.3\%, 95.45\% and 99.73\% confidence, are shown in blue, purple, and red, respectively.}
 \label{fig:phase-res-xspec}
\end{figure*}

For the phase-averaged data, the results of the spectral fitting, including the best-fit model, are shown in Figure~\ref{fig:xspec} and the best-fit model parameters are found in Table~\ref{table:best-fit}.
The \texttt{steppar} command in {\sc xspec} was used to create the confidence contours for the polarization measurements, and the resulting contour plots at 68.3\%, 95.45\% and 99.73\% confidence levels are presented in Figure~\ref{fig:phase-ave-contour}.
The results of the phase-averaged polarimetric analysis for the two different approaches give compatible results.

A wavy structure of the Stokes $U$ parameter residuals in Figure~\ref{fig:xspec} (see also Fig.~\ref{fig:phase-ave-PCUBE}) indicates that polarization is energy dependent. 
Thus, we replaced the \texttt{polconst} polarization model with the \texttt{pollin} model, corresponding to a linear energy dependence of the PD and PA. 
We assumed an energy-independent PA ($\psi_1$ in \textsc{xspec}; we fixed $\psi_{\rm slope}=0$), and allowed the PD to vary with photon energy $E$ (keV) as $\mbox{PD}(E)=A_1+A_{\rm slope}(E-1)$.  
This results in an improved fit ($\chi^2/\mathrm{d.o.f.}=2608.51/2221$ with an F-test probability of $2.3\times10^{-7}$). 
The best-fit parameters are $A_1=-5.3\pm2.3$\% and $A_{\rm slope}=2.2\pm0.6$\%, and PA=$\psi_1=-47\fdg1\pm6\fdg1$. 
The negative $A_1$ means that the PA at lower energies is rotated by 90\degr\ relative to the PA at higher energies and the PD is zero at $\approx$3.4 keV. 
Such a model is able to describe the observed behavior of Stokes $Q$ and $U$ parameters in Figures~\ref{fig:phase-ave-PCUBE} and \ref{fig:xspec}.   

For the phase-resolved spectro-polarimetric analysis, $I$, $Q$, and $U$ Stokes spectra were extracted for each phase bin individually, again utilizing the \texttt{xpbin} tool’s \texttt{PHA1}, \texttt{PHA1Q}, and \texttt{PHA1U} algorithms. 
The $I$, $Q$, and $U$ Stokes spectra were fitted with the same model as utilized for the phase-averaged spectro-polarimetric analysis, with the cross-calibration constants for DU2 and DU3 fixed to the values obtained for the phase-averaged analysis (see Table~\ref{table:best-fit}).
The results of the phase-resolved spectro-polarimetric analysis are summarized in Table ~\ref{table:phbins} and confidence contours corresponding to each phase bin are shown in Figure~\ref{fig:phase-res-xspec}.

A spectro-polarimetric phase-averaged analysis done separately for the eclipse data did not find a significant polarization, with an upper limit to the PD of 25.1\% at 99.73\% confidence level. 
An intensity-resolved spectro-polarimetric analysis performed on the combined data did not reveal a significant difference in polarization properties between different luminosity levels.
\clearpage 

\section{Discussion and summary}
\label{sec:sum}

XRPs are prime targets for X-ray polarimetric missions, where a high PD has been theoretically predicted because of the strong dependence of the primary radiation processes on the polarization of X-ray photons.
The birefringence of highly magnetized plasma allows for the radiative transfer to be treated in terms of two normal polarization modes: the ordinary (“O”) and extraordinary (“X”) mode \citep{1974JETP...38..903G}.
These two modes have different orientations in relation to the plane made up by the direction of the magnetic field and the momentum of photons.
For O-mode photons, the electric vector oscillates mainly within the plane, while for X-mode photons the electric vector oscillations are mainly oriented perpendicular to the plane.
Below the cyclotron energy the opacities of the polarization modes differ significantly, where the opacity of the X-mode is largely reduced compared to that of the O-mode \citep{2003PhRvL..91g1101L}, resulting in the predictions for PD as high as 80\% \citep{1988ApJ...324.1056M,2021MNRAS.501..109C}.

However, a significantly lower PD of only $\sim$2.3\% detected for \vela in the {IXPE} data is in line with recent measurements done for other accreting XRPs (Her X-1, \citealt{Doroshenko2022}; Cen X-3, \citealt{Tsygankov2022}; 4U~1626--67, \citealt{Marshall2022}), where similar, relatively low PDs have been reported.
The X-ray polarization of accreting XRPs greatly depends on the structure of the emission region (which is unknown), and most theoretical models mentioned here do not account yet for the temperature structure of the NS atmosphere,  assuming instead a uniform temperature in the spectral forming region.
In the case of the above-mentioned XRPs, the contradiction between the observed and the theoretically predicted values of the PD was explained with a model of the neutron star atmosphere overheated by the accretion process.  

The key factor for the depolarized emission in this model is a conversion of modes at the so-called vacuum resonance.
In strong magnetic fields, both the plasma and the vacuum are birefringent, where vacuum birefringence is a fundamental QED effect.
Generally speaking, the two effects (plasma and vacuum birefringence) tend to work against each other, and at vacuum resonance they cancel out \citep{1978PhRvL..41.1544M}. This leads to a transformation of the normal modes of radiation and a loss of the linear polarization degree.
The vacuum resonance occurs at a plasma density $\rho_{\mathrm{V}}\approx 10^{-4}B_{12}^{2}E_{\mathrm{keV}}^{2}$ $\mathrm{[g\;cm^{-3}]}$ for a given photon energy $E$ (in keV) and local magnetic field strength $B_{12}=B/10^{12}$ G.
This may result in a much smaller PD than normally predicted when considering the specific temperature structure of the atmospheres of accreting NSs. 
\citet{Doroshenko2022} have shown that a low PD of the X-ray radiation can be achieved if the point of vacuum resonance is located in an atmospheric transition layer with a strong temperature gradient.
If the transition region is located at the border of the overheated upper atmospheric layer and the cooler underlying atmosphere, the low PD occurs as a result of the fast mode conversion \citep{1978SvAL....4..117G}. 
For the specific atmospheric thickness of $\sim$3 $\mathrm{g\;cm^{-2}}$, corresponding to the Thomson optical depth around unity, a PD of the order of 10\% can be achieved \citep{Doroshenko2022}. 
As lower PDs than previously predicted seem to be a staple of sub-critical XRPs, theoretical models may well have to take into account the specific temperature structure of the NS atmosphere. 
On the other hand, the average observed luminosity for \vela is $\sim4\times10^{35}\;\mathrm{erg\;s^{-1}}$, roughly two orders of magnitude lower than the observed luminosities of Her~X-1 and Cen~X-3. 
Thus, it is rather puzzling that this scenario suggests a similar thickness of the overheated layer for a much smaller accretion rate; however, the key quantity here may be the proton stopping depth  \citep{1969SvA....13..175Z,1993ApJ...418..874N,2000ApJ...537..387Z,2019MNRAS.483..599G}.

The low PD may also be a result of the strong variations of the PD and PA with energy (see Figure~\ref{fig:phase-ave-PCUBE}), considering the evident $\sim90\degr$ difference in the PA below and above 3.5 keV. 
However, a complete and detailed analysis of the complicated energy dependence of the polarization properties of \vela is out of the scope of this paper and is subject to future, more extensive work.

Finally, we can speculate that the observed small PD is a result of strong variations of the PA with the pulsar phase. 
The observed pulse profile has a very complicated shape, which is related either to the complex structure of the surface magnetic field, or to the presence of a number of different components \citep[see][for discussion]{Tsygankov2022}.
The present photon statistics allowed us to obtain significant detection of polarization in 9 out of 12 phase bins, while to resolve the variations of the PA we likely needed many more bins.

\begin{acknowledgments}
The Imaging X-ray Polarimetry Explorer (IXPE) is a joint US and Italian mission.  The US contribution is supported by the National Aeronautics and Space Administration (NASA) and led and managed by its Marshall Space Flight Center (MSFC), with industry partner Ball Aerospace (contract NNM15AA18C).  
The Italian contribution is supported by the Italian Space Agency (Agenzia Spaziale Italiana, ASI) through contract ASI-OHBI-2017-12-I.0, agreements ASI-INAF-2017-12-H0 and ASI-INFN-2017.13-H0, and its Space
Science Data Center (SSDC) with agreements ASI-INAF-2022-14-HH.0 and
ASI-INFN 2021-43-HH.0, and by the Istituto Nazionale di Astrofisica
(INAF) and the Istituto Nazionale di Fisica Nucleare (INFN) in Italy.
This research used data products provided by the IXPE Team (MSFC, SSDC, INAF, and INFN) and distributed with additional software tools by the High-Energy Astrophysics Science Archive Research Center (HEASARC), at NASA Goddard Space Flight Center (GSFC).

We acknowledge support from the RSF grant 19-12-00423 (SST), the Academy of Finland grants 333112, 349144, 349373, and 349906 (JP, SST), the German Academic Exchange Service (DAAD) travel grant 57525212 (VD, VFS), and the German Research Foundation (DFG) grant WE 1312/53-1 (VFS). 
\end{acknowledgments}

%

\vspace{5mm}
\facilities{\ixpe, \swift (XRT and UVOT), \nustar}


\software{astropy \citep{2013A&A...558A..33A,2018AJ....156..123A},  
          {\sc xspec} \citep{Arn96}, 
          {\sc ixpeobssim} \citep{Baldini2022}.
          }






\begin{thebibliography}{}
\expandafter\ifx\csname natexlab\endcsname\relax\def\natexlab#1{#1}\fi
\providecommand{\url}[1]{\href{#1}{#1}}
\providecommand{\dodoi}[1]{doi:~\href{http://doi.org/#1}{\nolinkurl{#1}}}
\providecommand{\doeprint}[1]{\href{http://ascl.net/#1}{\nolinkurl{http://ascl.net/#1}}}
\providecommand{\doarXiv}[1]{\href{https://arxiv.org/abs/#1}{\nolinkurl{https://arxiv.org/abs/#1}}}

\bibitem[{{Arnaud}(1996)}]{Arn96}
{Arnaud}, K.~A. 1996, in ASP Conf. Ser., Vol. 101, Astronomical Data Analysis
  Software and Systems V, ed. G.~H. {Jacoby} \& J.~{Barnes} (San Francisco:
  Astron. Soc. Pac.), 17--20

\bibitem[{{Astropy Collaboration} {et~al.}(2013){Astropy Collaboration},
  {Robitaille}, {Tollerud}, {Greenfield}, {Droettboom}, {Bray}, {Aldcroft},
  {Davis}, {Ginsburg}, {Price-Whelan}, {Kerzendorf}, {Conley}, {Crighton},
  {Barbary}, {Muna}, {Ferguson}, {Grollier}, {Parikh}, {Nair}, {Unther},
  {Deil}, {Woillez}, {Conseil}, {Kramer}, {Turner}, {Singer}, {Fox}, {Weaver},
  {Zabalza}, {Edwards}, {Azalee Bostroem}, {Burke}, {Casey}, {Crawford},
  {Dencheva}, {Ely}, {Jenness}, {Labrie}, {Lim}, {Pierfederici}, {Pontzen},
  {Ptak}, {Refsdal}, {Servillat}, \& {Streicher}}]{2013A&A...558A..33A}
{Astropy Collaboration}, {Robitaille}, T.~P., {Tollerud}, E.~J., {et~al.} 2013,
  \aap, 558, A33, \dodoi{10.1051/0004-6361/201322068}

\bibitem[{{Astropy Collaboration} {et~al.}(2018){Astropy Collaboration},
  {Price-Whelan}, {Sip{\H{o}}cz}, {G{\"u}nther}, {Lim}, {Crawford}, {Conseil},
  {Shupe}, {Craig}, {Dencheva}, {Ginsburg}, {VanderPlas}, {Bradley},
  {P{\'e}rez-Su{\'a}rez}, {de Val-Borro}, {Aldcroft}, {Cruz}, {Robitaille},
  {Tollerud}, {Ardelean}, {Babej}, {Bach}, {Bachetti}, {Bakanov}, {Bamford},
  {Barentsen}, {Barmby}, {Baumbach}, {Berry}, {Biscani}, {Boquien}, {Bostroem},
  {Bouma}, {Brammer}, {Bray}, {Breytenbach}, {Buddelmeijer}, {Burke},
  {Calderone}, {Cano Rodr{\'\i}guez}, {Cara}, {Cardoso}, {Cheedella}, {Copin},
  {Corrales}, {Crichton}, {D'Avella}, {Deil}, {Depagne}, {Dietrich}, {Donath},
  {Droettboom}, {Earl}, {Erben}, {Fabbro}, {Ferreira}, {Finethy}, {Fox},
  {Garrison}, {Gibbons}, {Goldstein}, {Gommers}, {Greco}, {Greenfield},
  {Groener}, {Grollier}, {Hagen}, {Hirst}, {Homeier}, {Horton}, {Hosseinzadeh},
  {Hu}, {Hunkeler}, {Ivezi{\'c}}, {Jain}, {Jenness}, {Kanarek}, {Kendrew},
  {Kern}, {Kerzendorf}, {Khvalko}, {King}, {Kirkby}, {Kulkarni}, {Kumar},
  {Lee}, {Lenz}, {Littlefair}, {Ma}, {Macleod}, {Mastropietro}, {McCully},
  {Montagnac}, {Morris}, {Mueller}, {Mumford}, {Muna}, {Murphy}, {Nelson},
  {Nguyen}, {Ninan}, {N{\"o}the}, {Ogaz}, {Oh}, {Parejko}, {Parley}, {Pascual},
  {Patil}, {Patil}, {Plunkett}, {Prochaska}, {Rastogi}, {Reddy Janga},
  {Sabater}, {Sakurikar}, {Seifert}, {Sherbert}, {Sherwood-Taylor}, {Shih},
  {Sick}, {Silbiger}, {Singanamalla}, {Singer}, {Sladen}, {Sooley},
  {Sornarajah}, {Streicher}, {Teuben}, {Thomas}, {Tremblay}, {Turner},
  {Terr{\'o}n}, {van Kerkwijk}, {de la Vega}, {Watkins}, {Weaver}, {Whitmore},
  {Woillez}, {Zabalza}, \& {Astropy Contributors}}]{2018AJ....156..123A}
{Astropy Collaboration}, {Price-Whelan}, A.~M., {Sip{\H{o}}cz}, B.~M., {et~al.}
  2018, \aj, 156, 123, \dodoi{10.3847/1538-3881/aabc4f}

\bibitem[{{Baldini} {et~al.}(2022){Baldini}, {Bucciantini}, {Di Lalla},
  {Ehlert}, {Manfreda}, {Omodei}, {Pesce-Rollins}, \& {Sgr{\`o}}}]{Baldini2022}
{Baldini}, L., {Bucciantini}, N., {Di Lalla}, N., {et~al.} 2022, SoftwareX, 19,
  101194, \dodoi{10.1016/j.softx.2022.101194}

\bibitem[{{Caiazzo} \& {Heyl}(2021)}]{2021MNRAS.501..109C}
{Caiazzo}, I., \& {Heyl}, J. 2021, \mnras, 501, 109,
  \dodoi{10.1093/mnras/staa3428}

\bibitem[{{Chodil} {et~al.}(1967){Chodil}, {Mark}, {Rodrigues}, {Seward}, \&
  {Swift}}]{1967ApJ...150...57C}
{Chodil}, G., {Mark}, H., {Rodrigues}, R., {Seward}, F.~D., \& {Swift}, C.~D.
  1967, \apj, 150, 57, \dodoi{10.1086/149312}

\bibitem[{{Di Marco} {et~al.}(2023){Di Marco}, {Soffitta}, {Costa},
  {Ferrazzoli}, {La Monaca}, {Rankin}, {Ratheesh}, {Xie}, {Baldini}, {Del
  Monte}, {Ehlert}, {Fabiani}, {Kim}, {Muleri}, {O'Dell}, {Ramsey}, {Rubini},
  {Sgr{\`o}}, {Silvestri}, {Tennant}, \& {Weisskopf}}]{2023-IXPE-background}
{Di Marco}, A., {Soffitta}, P., {Costa}, E., {et~al.} 2023, arXiv e-prints,
  arXiv:2302.02927, \dodoi{10.48550/arXiv.2302.02927}

\bibitem[{{Doroshenko} {et~al.}(2022){Doroshenko}, {Poutanen}, {Tsygankov},
  {Suleimanov}, {Bachetti}, {Caiazzo}, {Costa}, {Di Marco}, {Heyl}, {La
  Monaca}, {Muleri}, {Mushtukov}, {Pavlov}, {Ramsey}, {Rankin}, {Santangelo},
  {Soffitta}, {Staubert}, {Weisskopf}, {Zane}, {Agudo}, {Antonelli}, {Baldini},
  {Baumgartner}, {Bellazzini}, {Bianchi}, {Bongiorno}, {Bonino}, {Brez},
  {Bucciantini}, {Capitanio}, {Castellano}, {Cavazzuti}, {Ciprini}, {De Rosa},
  {Del Monte}, {Di Gesu}, {Di Lalla}, {Donnarumma}, {Dov{\v{c}}iak}, {Ehlert},
  {Enoto}, {Evangelista}, {Fabiani}, {Ferrazzoli}, {Garcia}, {Gunji},
  {Hayashida}, {Iwakiri}, {Jorstad}, {Karas}, {Kitaguchi}, {Kolodziejczak},
  {Krawczynski}, {Latronico}, {Liodakis}, {Maldera}, {Manfreda}, {Marin},
  {Marinucci}, {Marscher}, {Marshall}, {Matt}, {Mitsuishi}, {Mizuno}, {Ng},
  {O'Dell}, {Omodei}, {Oppedisano}, {Papitto}, {Peirson}, {Perri},
  {Pesce-Rollins}, {Pilia}, {Possenti}, {Puccetti}, {Ratheesh}, {Romani},
  {Sgr{\`o}}, {Slane}, {Spandre}, {Sunyaev}, {Tamagawa}, {Tavecchio},
  {Taverna}, {Tawara}, {Tennant}, {Thomas}, {Tombesi}, {Trois}, {Turolla},
  {Vink}, {Wu}, \& {Xie}}]{Doroshenko2022}
{Doroshenko}, V., {Poutanen}, J., {Tsygankov}, S.~S., {et~al.} 2022, Nature
  Astronomy, 6, 1433, \dodoi{10.1038/s41550-022-01799-5}

\bibitem[{{F{\"u}rst} {et~al.}(2014){F{\"u}rst}, {Pottschmidt}, {Wilms},
  {Tomsick}, {Bachetti}, {Boggs}, {Christensen}, {Craig}, {Grefenstette},
  {Hailey}, {Harrison}, {Madsen}, {Miller}, {Stern}, {Walton}, \&
  {Zhang}}]{2014ApJ...780..133F}
{F{\"u}rst}, F., {Pottschmidt}, K., {Wilms}, J., {et~al.} 2014, \apj, 780, 133,
  \dodoi{10.1088/0004-637X/780/2/133}

\bibitem[{{Gnedin} \& {Pavlov}(1974)}]{1974JETP...38..903G}
{Gnedin}, Y.~N., \& {Pavlov}, G.~G. 1974, Soviet Journal of Experimental and
  Theoretical Physics, 38, 903

\bibitem[{{Gnedin} {et~al.}(1978){Gnedin}, {Pavlov}, \&
  {Shibanov}}]{1978SvAL....4..117G}
{Gnedin}, Y.~N., {Pavlov}, G.~G., \& {Shibanov}, Y.~A. 1978, Soviet Astronomy
  Letters, 4, 117

\bibitem[{{Gonz{\'a}lez-Caniulef} {et~al.}(2019){Gonz{\'a}lez-Caniulef},
  {Zane}, {Turolla}, \& {Wu}}]{2019MNRAS.483..599G}
{Gonz{\'a}lez-Caniulef}, D., {Zane}, S., {Turolla}, R., \& {Wu}, K. 2019,
  \mnras, 483, 599, \dodoi{10.1093/mnras/sty3159}

\bibitem[{{Ji} {et~al.}(2019){Ji}, {Staubert}, {Ducci}, {Santangelo}, {Zhang},
  \& {Chang}}]{2019MNRAS.484.3797J}
{Ji}, L., {Staubert}, R., {Ducci}, L., {et~al.} 2019, \mnras, 484, 3797,
  \dodoi{10.1093/mnras/stz264}

\bibitem[{{Kendziorra} {et~al.}(1992){Kendziorra}, {Mony}, {Kretschmar},
  {Maisack}, {Staubert}, {D{\"o}bereiner}, {Englhauser}, {Pietsch}, {Reppin},
  {Tr{\"u}mper}, {Efremov}, {Kaniovsky}, \& {Sunyaev}}]{1992fxra.conf...51K}
{Kendziorra}, E., {Mony}, B., {Kretschmar}, P., {et~al.} 1992, in Proc. Yamada
  Conf. XXVIII, Frontiers Science Series, ed. Y.~{Tanaka} \& K.~{Koyama}
  (Tokyo: Universal Academy Press), 51

\bibitem[{{Kislat} {et~al.}(2015){Kislat}, {Clark}, {Beilicke}, \&
  {Krawczynski}}]{2015-Kislat}
{Kislat}, F., {Clark}, B., {Beilicke}, M., \& {Krawczynski}, H. 2015,
  Astroparticle Physics, 68, 45, \dodoi{10.1016/j.astropartphys.2015.02.007}

\bibitem[{{Kretschmar} {et~al.}(1996){Kretschmar}, {Pan}, {Kendziorra}, {Kunz},
  {Maisack}, {Staubert}, {Pietsch}, {Truemper}, {Efremov}, \&
  {Sunyaev}}]{1996A&AS..120C.175K}
{Kretschmar}, P., {Pan}, H.~C., {Kendziorra}, E., {et~al.} 1996, \aaps, 120,
  175

\bibitem[{{Kretschmar} {et~al.}(1997){Kretschmar}, {Pan}, {Kendziorra},
  {Maisack}, {Staubert}, {Skinner}, {Pietsch}, {Truemper}, {Efremov}, \&
  {Sunyaev}}]{1997A&A...325..623K}
---. 1997, \aap, 325, 623

\bibitem[{{Kretschmar} {et~al.}(2021){Kretschmar}, {El Mellah},
  {Mart{\'\i}nez-N{\'u}{\~n}ez}, {F{\"u}rst}, {Grinberg}, {Sander}, {van den
  Eijnden}, {Degenaar}, {Ma{\'\i}z Apell{\'a}niz}, {Jim{\'e}nez Esteban},
  {Ramos-Lerate}, \& {Utrilla}}]{2021A&A...652A..95K}
{Kretschmar}, P., {El Mellah}, I., {Mart{\'\i}nez-N{\'u}{\~n}ez}, S., {et~al.}
  2021, \aap, 652, A95, \dodoi{10.1051/0004-6361/202040272}

\bibitem[{{Kreykenbohm} {et~al.}(2008){Kreykenbohm}, {Wilms}, {Kretschmar},
  {Torrej{\'o}n}, {Pottschmidt}, {Hanke}, {Santangelo}, {Ferrigno}, \&
  {Staubert}}]{2008A&A...492..511K}
{Kreykenbohm}, I., {Wilms}, J., {Kretschmar}, P., {et~al.} 2008, \aap, 492,
  511, \dodoi{10.1051/0004-6361:200809956}

\bibitem[{{La Parola} {et~al.}(2016){La Parola}, {Cusumano}, {Segreto}, \&
  {D'A{\`\i}}}]{2016MNRAS.463..185L}
{La Parola}, V., {Cusumano}, G., {Segreto}, A., \& {D'A{\`\i}}, A. 2016,
  \mnras, 463, 185, \dodoi{10.1093/mnras/stw1915}

\bibitem[{{Lai} \& {Ho}(2003)}]{2003PhRvL..91g1101L}
{Lai}, D., \& {Ho}, W.~C. 2003, \prl, 91, 071101,
  \dodoi{10.1103/PhysRevLett.91.071101}

\bibitem[{{Maier} {et~al.}(2014){Maier}, {Tenzer}, \& {Santangelo}}]{Maier2014}
{Maier}, D., {Tenzer}, C., \& {Santangelo}, A. 2014, \pasp, 126, 459,
  \dodoi{10.1086/676820}

\bibitem[{{Makishima} \& {Mihara}(1992)}]{1992fxra.conf...23M}
{Makishima}, K., \& {Mihara}, T. 1992, in Proc. Yamada Conf. XXVIII, Frontiers
  Science Series, ed. Y.~{Tanaka} \& K.~{Koyama} (Tokyo: Universal Academy
  Press), 23

\bibitem[{{Marshall} {et~al.}(2022){Marshall}, {Ng}, {Rogantini}, {Heyl},
  {Tsygankov}, {Poutanen}, {Costa}, {Zane}, {Malacaria}, {Agudo}, {Antonelli},
  {Bachetti}, {Baldini}, {Baumgartner}, {Bellazzini}, {Bianchi}, {Bongiorno},
  {Bonino}, {Brez}, {Bucciantini}, {Capitanio}, {Castellano}, {Cavazzuti},
  {Ciprini}, {De Rosa}, {Del Monte}, {Di Gesu}, {Di Lalla}, {Di Marco},
  {Donnarumma}, {Doroshenko}, {Dovvciak}, {Ehlert}, {Enoto}, {Evangelista},
  {Fabiani}, {Ferrazzoli}, {Garcia}, {Gunji}, {Hayashida}, {Iwakiri},
  {Jorstad}, {Karas}, {Kitaguchi}, {Kolodziejczak}, {Krawczynski}, {La Monaca},
  {Latronico}, {Liodakis}, {Maldera}, {Manfreda}, {Marin}, {Marinucci},
  {Marscher}, {Matt}, {Mitsuishi}, {Mizuno}, {Muleri}, {Ng}, {ODell}, {Omodei},
  {Oppedisano}, {Papitto}, {Pavlov}, {Peirson}, {Perri}, {Pesce-Rollins},
  {Petrucci}, {Pilia}, {Possenti}, {Puccetti}, {Ramsey}, {Rankin}, {Ratheesh},
  {Romani}, {Sgro}, {Slane}, {Soffitta}, {Spandre}, {Tamagawa}, {Tavecchio},
  {Taverna}, {Tawara}, {Tennant}, {Thomas}, {Tombesi}, {Trois}, {Turolla},
  {Vink}, {Weisskopf}, {Wu}, {Xie}, {Schulz}, \& {Chakrabarty}}]{Marshall2022}
{Marshall}, H.~L., {Ng}, M., {Rogantini}, D., {et~al.} 2022, \apj, 940, 70,
  \dodoi{10.3847/1538-4357/ac98c2}

\bibitem[{{McClintock} {et~al.}(1976){McClintock}, {Rappaport}, {Joss},
  {Bradt}, {Buff}, {Clark}, {Hearn}, {Lewin}, {Matilsky}, {Mayer}, \&
  {Primini}}]{1976ApJ...206L..99M}
{McClintock}, J.~E., {Rappaport}, S., {Joss}, P.~C., {et~al.} 1976, \apjl, 206,
  L99, \dodoi{10.1086/182142}

\bibitem[{{Meszaros} {et~al.}(1988){Meszaros}, {Novick}, {Szentgyorgyi},
  {Chanan}, \& {Weisskopf}}]{1988ApJ...324.1056M}
{Meszaros}, P., {Novick}, R., {Szentgyorgyi}, A., {Chanan}, G.~A., \&
  {Weisskopf}, M.~C. 1988, \apj, 324, 1056, \dodoi{10.1086/165962}

\bibitem[{{M{\'e}sz{\'a}ros} \& {Ventura}(1978)}]{1978PhRvL..41.1544M}
{M{\'e}sz{\'a}ros}, P., \& {Ventura}, J. 1978, \prl, 41, 1544,
  \dodoi{10.1103/PhysRevLett.41.1544}

\bibitem[{{Mikhalev}(2018)}]{Mikhalev2018}
{Mikhalev}, V. 2018, \aap, 615, A54, \dodoi{10.1051/0004-6361/201731971}

\bibitem[{{Mushtukov} \& {Tsygankov}(2022)}]{2022arXiv220414185M}
{Mushtukov}, A., \& {Tsygankov}, S. 2022, arXiv e-prints, arXiv:2204.14185.
\newblock \doarXiv{2204.14185}

\bibitem[{{Nagase} {et~al.}(1986){Nagase}, {Hayakawa}, {Sato}, {Masai}, \&
  {Inoue}}]{1986PASJ...38..547N}
{Nagase}, F., {Hayakawa}, S., {Sato}, N., {Masai}, K., \& {Inoue}, H. 1986,
  \pasj, 38, 547

\bibitem[{{Nelson} {et~al.}(1993){Nelson}, {Salpeter}, \&
  {Wasserman}}]{1993ApJ...418..874N}
{Nelson}, R.~W., {Salpeter}, E.~E., \& {Wasserman}, I. 1993, \apj, 418, 874,
  \dodoi{10.1086/173445}

\bibitem[{{Quaintrell} {et~al.}(2003){Quaintrell}, {Norton}, {Ash}, {Roche},
  {Willems}, {Bedding}, {Baldry}, \& {Fender}}]{2003A&A...401..313Q}
{Quaintrell}, H., {Norton}, A.~J., {Ash}, T.~D.~C., {et~al.} 2003, \aap, 401,
  313, \dodoi{10.1051/0004-6361:20030120}

\bibitem[{{Serkowski}(1958)}]{Serkowski1958}
{Serkowski}, K. 1958, \actaa, 8, 135

\bibitem[{{Simmons} \& {Stewart}(1985)}]{Simmons1985}
{Simmons}, J.~F.~L., \& {Stewart}, B.~G. 1985, \aap, 142, 100

\bibitem[{{Soffitta} {et~al.}(2021){Soffitta}, {Baldini}, {Bellazzini},
  {Costa}, {Latronico}, {Muleri}, {Del Monte}, {Fabiani}, {Minuti}, {Pinchera},
  {Sgro'}, {Spandre}, {Trois}, {Amici}, {Andersson}, {Attina'}, {Bachetti},
  {Barbanera}, {Borotto}, {Brez}, {Brienza}, {Caporale}, {Cardelli},
  {Carpentiero}, {Castellano}, {Castronuovo}, {Cavalli}, {Cavazzuti},
  {Ceccanti}, {Centrone}, {Ciprini}, {Citraro}, {D'Amico}, {D'Alba}, {Di
  Cosimo}, {Di Lalla}, {Di Marco}, {Di Persio}, {Donnarumma}, {Evangelista},
  {Ferrazzoli}, {Hayato}, {Kitaguchi}, {La Monaca}, {Lefevre}, {Loffredo},
  {Lorenzi}, {Lucchesi}, {Magazzu}, {Maldera}, {Manfreda}, {Mangraviti},
  {Marengo}, {Matt}, {Mereu}, {Morbidini}, {Mosti}, {Nakano}, {Nasimi},
  {Negri}, {Nenonen}, {Nuti}, {Orsini}, {Perri}, {Pesce-Rollins}, {Piazzolla},
  {Pilia}, {Profeti}, {Puccetti}, {Rankin}, {Ratheesh}, {Rubini}, {Santoli},
  {Sarra}, {Scalise}, {Sciortino}, {Tamagawa}, {Tardiola}, {Tobia},
  {Vimercati}, \& {Xie}}]{Soffitta21}
{Soffitta}, P., {Baldini}, L., {Bellazzini}, R., {et~al.} 2021, \aj, 162, 208,
  \dodoi{10.3847/1538-3881/ac19b0}

\bibitem[{{Staubert} {et~al.}(2004){Staubert}, {Kreykenbohm}, {Kretschmar},
  {Chernyakova}, {Pottschmidt}, {Benlloch-Garcia}, {Wilms}, {Santangelo},
  {Segreto}, {von Kienlin}, {Sidoli}, {Larsson}, \&
  {Westergaard}}]{2004ESASP.552..259S}
{Staubert}, R., {Kreykenbohm}, I., {Kretschmar}, P., {et~al.} 2004, in ESA
  Spec. Publ., Vol. 552, 5th INTEGRAL Workshop, the INTEGRAL Universe, ed.
  V.~{Schoenfelder}, G.~{Lichti}, \& C.~{Winkler}, 259

\bibitem[{{Tsygankov} {et~al.}(2022){Tsygankov}, {Doroshenko}, {Poutanen},
  {Heyl}, {Mushtukov}, {Caiazzo}, {Di Marco}, {Forsblom},
  {Gonz{\'a}lez-Caniulef}, {Klawin}, {La Monaca}, {Malacaria}, {Marshall},
  {Muleri}, {Ng}, {Suleimanov}, {Sunyaev}, {Turolla}, {Agudo}, {Antonelli},
  {Bachetti}, {Baldini}, {Baumgartner}, {Bellazzini}, {Bianchi}, {Bongiorno},
  {Bonino}, {Brez}, {Bucciantini}, {Capitanio}, {Castellano}, {Cavazzuti},
  {Ciprini}, {Costa}, {De Rosa}, {Del Monte}, {Di Gesu}, {Di Lalla},
  {Donnarumma}, {Dov{\v{c}}iak}, {Ehlert}, {Enoto}, {Evangelista}, {Fabiani},
  {Ferrazzoli}, {Garcia}, {Gunji}, {Hayashida}, {Iwakiri}, {Jorstad}, {Karas},
  {Kitaguchi}, {Kolodziejczak}, {Krawczynski}, {Latronico}, {Liodakis},
  {Maldera}, {Manfreda}, {Marin}, {Marinucci}, {Marscher}, {Matt}, {Mitsuishi},
  {Mizuno}, {Ng}, {O'Dell}, {Omodei}, {Oppedisano}, {Papitto}, {Pavlov},
  {Peirson}, {Perri}, {Pesce-Rollins}, {Petrucci}, {Pilia}, {Possenti},
  {Puccetti}, {Ramsey}, {Rankin}, {Ratheesh}, {Romani}, {Sgr{\`o}}, {Slane},
  {Soffitta}, {Spandre}, {Tamagawa}, {Tavecchio}, {Taverna}, {Tawara},
  {Tennant}, {Thomas}, {Tombesi}, {Trois}, {Vink}, {Weisskopf}, {Wu}, {Xie},
  {Zane}, \& {IXPE Collaboration}}]{Tsygankov2022}
{Tsygankov}, S.~S., {Doroshenko}, V., {Poutanen}, J., {et~al.} 2022, \apjl,
  941, L14, \dodoi{10.3847/2041-8213/aca486}

\bibitem[{{Ulmer} {et~al.}(1972){Ulmer}, {Baity}, {Wheaton}, \&
  {Peterson}}]{1972ApJ...178L.121U}
{Ulmer}, M.~P., {Baity}, W.~A., {Wheaton}, W.~A., \& {Peterson}, L.~E. 1972,
  \apjl, 178, L121, \dodoi{10.1086/181099}

\bibitem[{{van Kerkwijk} {et~al.}(1995){van Kerkwijk}, {van Paradijs},
  {Zuiderwijk}, {Hammerschlag-Hensberge}, {Kaper}, \&
  {Sterken}}]{1995A&A...303..483V}
{van Kerkwijk}, M.~H., {van Paradijs}, J., {Zuiderwijk}, E.~J., {et~al.} 1995,
  \aap, 303, 483.
\newblock \doarXiv{astro-ph/9505070}

\bibitem[{{Weisskopf} {et~al.}(2010){Weisskopf}, {Elsner}, \&
  {O'Dell}}]{2010SPIE.7732E..0EW}
{Weisskopf}, M.~C., {Elsner}, R.~F., \& {O'Dell}, S.~L. 2010, in \procspie,
  Vol. 7732, Space Telescopes and Instrumentation 2010: Ultraviolet to Gamma
  Ray, ed. M.~{Arnaud}, S.~S. {Murray}, \& T.~{Takahashi}, 77320E,
  \dodoi{10.1117/12.857357}

\bibitem[{{Weisskopf} {et~al.}(2022){Weisskopf}, {Soffitta}, {Baldini},
  {Ramsey}, {O'Dell}, {Romani}, {Matt}, {Deininger}, {Baumgartner},
  {Bellazzini}, {Costa}, {Kolodziejczak}, {Latronico}, {Marshall}, {Muleri},
  {Bongiorno}, {Tennant}, {Bucciantini}, {Dovciak}, {Marin}, {Marscher},
  {Poutanen}, {Slane}, {Turolla}, {Kalinowski}, {Di Marco}, {Fabiani},
  {Minuti}, {La Monaca}, {Pinchera}, {Rankin}, {Sgro'}, {Trois}, {Xie},
  {Alexander}, {Allen}, {Amici}, {Andersen}, {Antonelli}, {Antoniak},
  {Attina'}, {Barbanera}, {Bachetti}, {Baggett}, {Bladt}, {Brez}, {Bonino},
  {Boree}, {Borotto}, {Breeding}, {Brienza}, {Bygott}, {Caporale}, {Cardelli},
  {Carpentiero}, {Castellano}, {Castronuovo}, {Cavalli}, {Cavazzuti},
  {Ceccanti}, {Centrone}, {Citraro}, {D' Amico}, {D'Alba}, {Di Gesu}, {Del
  Monte}, {Dietz}, {Di Lalla}, {Di Persio}, {Dolan}, {Donnarumma},
  {Evangelista}, {Ferrant}, {Ferrazzoli}, {Ferrie}, {Footdale}, {Forsyth},
  {Foster}, {Garelick}, {Gunji}, {Gurnee}, {Head}, {Hibbard}, {Johnson},
  {Kelly}, {Kilaru}, {Lefevre}, {Le Roy}, {Loffredo}, {Lorenzi}, {Lucchesi},
  {Maddox}, {Magazzu}, {Maldera}, {Manfreda}, {Mangraviti}, {Marengo},
  {Marrocchesi}, {Massaro}, {Mauger}, {McCracken}, {McEachen}, {Mize}, {Mereu},
  {Mitchell}, {Mitsuishi}, {Morbidini}, {Mosti}, {Nasimi}, {Negri}, {Negro},
  {Nguyen}, {Nitschke}, {Nuti}, {Onizuka}, {Oppedisano}, {Orsini}, {Osborne},
  {Pacheco}, {Paggi}, {Painter}, {Pavelitz}, {Pentz}, {Piazzolla}, {Perri},
  {Pesce-Rollins}, {Peterson}, {Pilia}, {Profeti}, {Puccetti}, {Ranganathan},
  {Ratheesh}, {Reedy}, {Root}, {Rubini}, {Ruswick}, {Sanchez}, {Sarra},
  {Santoli}, {Scalise}, {Sciortino}, {Schroeder}, {Seek}, {Sosdian}, {Spandre},
  {Speegle}, {Tamagawa}, {Tardiola}, {Tobia}, {Thomas}, {Valerie}, {Vimercati},
  {Walden}, {Weddendorf}, {Wedmore}, {Welch}, {Zanetti}, \&
  {Zanetti}}]{Weisskopf2022}
{Weisskopf}, M.~C., {Soffitta}, P., {Baldini}, L., {et~al.} 2022, J. Astron.
  Telesc. Instrum. Syst., 8, 026002, \dodoi{10.1117/1.JATIS.8.2.026002}

\bibitem[{{Wilms} {et~al.}(2000){Wilms}, {Allen}, \& {McCray}}]{Wilms2000}
{Wilms}, J., {Allen}, A., \& {McCray}, R. 2000, \apj, 542, 914,
  \dodoi{10.1086/317016}

\bibitem[{{Zane} {et~al.}(2000){Zane}, {Turolla}, \&
  {Treves}}]{2000ApJ...537..387Z}
{Zane}, S., {Turolla}, R., \& {Treves}, A. 2000, \apj, 537, 387,
  \dodoi{10.1086/309027}

\bibitem[{{Zel'dovich} \& {Shakura}(1969)}]{1969SvA....13..175Z}
{Zel'dovich}, Y.~B., \& {Shakura}, N.~I. 1969, \sovast, 13, 175

\end{thebibliography}



\end{document}